\documentclass[final,authoryear,5p]{elsarticle}

\usepackage{graphicx}
\usepackage{lineno}
\usepackage{amssymb}

\usepackage{amsmath}
\usepackage{calc}

\usepackage[ps2pdf,%
a4paper=true,%
breaklinks=true,%
colorlinks=true,%
pdfauthor={First Author et al.},%
pdftitle={Template for manuscripts in Advances in Space Research}%
]{hyperref}

\journal{Advances in Space Research}
\begin{document}

\begin{frontmatter}




\title{A study of VLF signals variations associated with the changes of ionization level in the D-region in consequence of solar conditions}

\author[label1]{D.M. {\v S}uli{\' c}}
\ead{desanka.sulic@gmail.com}
\author[label2]{V.A. Sre{\'c}kovi{\' c}}
\ead{vlada@ipb.ac.rs}
\author[label2]{A.A. Mihajlov}
\ead{mihajlov@ipb.ac.rs}
 \address[label1]{University Union - Nikola Tesla, 11000, Belgrade, Serbia}
 \address[label2]{University of Belgrade, Institute of Physics, P. O. Box 57, 11001 Belgrade, Serbia}

\begin{abstract}
In this paper we confine our attention to the analysis of amplitude and phase data acquired by monitoring VLF/LF radio signals emitted by four European transmitters during a seven-year period (2008-2014).  All the data were recorded at a Belgrade site (44.85$^{0}$ N, 20.38$^{0}$ E) by the Stanford University ELF/VLF receiver AWESOME. Propagation of VLF/LF radio signal takes place in the Earth-ionosphere waveguide and strongly depends on ionization level of the D-region, which means that it is mainly controlled by solar conditions. Some results of amplitude and phase variations on GQD/22.10 kHz, DHO/23.40 kHz, ICV/20.27 kHz and NSC/45.90 kHz radio signals measurements at short distances ($D < 2$ Mm) over Central Europe and their interpretation are summarized in this paper.  Attention is restricted to regular diurnal, seasonal and solar variations including sunrise and sunset effects on propagation characteristics of four VLF/LF radio signals. We study VLF/LF propagation over short path as a superposition of different number of discrete modes which depends on the variations of the path parameters. Although the solar X-ray flare effects on propagation of VLF/LF radio signals are well recognized on all paths, similarities and differences between them are defined under existing conditions over the paths. Statistical results show that the size of amplitude and phase perturbations on VLF/LF radio signal is in correlation with the intensity of X-ray flux. We present the calculations of electron density enhancements in the D-region caused by different classes of solar X-ray flares during the period of ascending phase and maximum of the solar cycle 24.
\end{abstract}

\begin{keyword}
Solar activity \sep Solar flare response \sep D-region \sep VLF  \sep Ionospheric disturbances
\PACS 96.60.qe \sep 96.60.qd \sep 96.60.Q- \sep 92.60.-e \sep 94.20.-y \sep 41.20.Jb
\end{keyword}

\end{frontmatter}

\parindent=0.5 cm

\section{Introduction}

The lowest region of the ionosphere, the D-region, is important as a reflecting layer for the longwave communication and navigation systems. The \underbar{V}ery \underbar{L}ow \underbar{F}requency (VLF, 3-30 kHz) and \underbar{L}ow \underbar{F}requency (LF, 30-300 kHz) bands are below the critical frequencies of the D-region. VLF/LF radio waves from transmitters propagate through waveguide bounded by the Earth's surface and the D-region. This propagation is stable both in amplitude and phase and has relatively low attenuation. VLF/LF radiation tends to reflect from electron densities (strictly conductivities) at altitudes of 70-75 km during daytime and 80-90 km during nighttime. Also VLF/LF radiation is reflected by the conducting Earth's surface and this means that the radio waves propagate over Earth trapped between the imperfect mirrors of the ground and the ionosphere \citep{wai64,mit74}. The effective reflection height depends on the ionization levels of the D-region. The lowest region of the ionosphere ($<$ 90 km altitude) is formed during quiet conditions primarily by the action of solar Lyman-$\alpha$ radiation (121.6 nm) on nitric oxide. Daytime electron density in this region is about or less than $Ne \sim 10^{8}$ m$^{-3}$. During the nighttime the ionization rate drops and recombination processes continue. Even at night there is a sufficient ionization in the lowest region of ionosphere to affect VLF/LF radio signals \citep{goo05, kel09}.


A range of dynamic phenomena occur in the D-region and cause diurnal and seasonal variations in connection with solar activity (11-year sunspot cycle). The phenomenon such as solar X-ray flare illuminating the daytime ionosphere induces unpredictable effects that are associated with space weather. When the solar X-ray flares appear, the X-ray fluxes suddenly increase and the ones with the appreciable wavelength below 1 nm are able to penetrate down to the D-region and increase the ionization rate there \citep{tho01}. A lot of work has been done regarding the correlation between X-ray fluxes and VLF perturbations as well as D-region electron density profile \citep{tho93,tho01,zig07}. The changes in the conditions of the D-region at these altitudes cause the changes in the received amplitude and phase at the receiver, allowing us to compare experimental observations of received radio signals with the simulations based upon the predicted changes in the D-region to understand what is happening.

\section{Data analysis method}

\subsection{Description of experimental data}  In this paper we confine our attention to the analysis of amplitude and phase data acquired by monitoring VLF/LF radio signals emitted by four European transmitters during a seven-year period (2008-2014). This period covers the ascending phase and maximum of the solar cycle 24. All the data were recorded at a Belgrade site (44.85$^{0}$ N, 20.38$^{0}$ E), Serbia by the Stanford University ELF/VLF Receiver \underbar{A}tmospheric \underbar{W}eather \underbar{E}lectromagnetic \underbar{S}ystem for \underbar{O}bservation \underbar{M}odeling and \underbar{E}ducation (AWESOME). Narrowband data can be recorded in a continuous fashion, even in case when as many as 15 transmitters are being monitored \citep{coh10}.

VLF/LF radio signals received at Belgrade site include: GQD/22.10 kHz, DHO/23.40 kHz, ICV/20.27 kHz and NSC/45.90 kHz. These VLF/LF radio signals propagate in the Earth-ionosphere waveguides over Central Europe. \underline{G}reat \underline{C}ircle \underline{P}aths (GCPs) for those signals are short and we divide them in two groups: $D$ $<$ 1Mm and 1 Mm $<$ $D$ $<$ 2 Mm. The transmitters of radio signals: DHO/23.40 kHz, ICV/20.27 kHz and NSC/45.90 kHz are located in the same time zone (local time: UT+1 hour) as a receiver site. The transmitter of GQD/22.10 kHz is located in the UK, but a radio signal propagates over great segment of path in the same time zone where the receiver is.

\begin{figure}[h]
\begin{center}
\includegraphics[width=0.75\columnwidth,height=0.45\columnwidth]{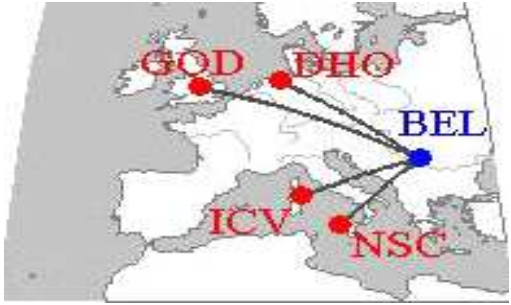} \caption{\underline{G}reat \underline{C}ircle \underline{P}aths (GCPs) of subionospherically propagating VLF/LF radio signals recorded at Belgrade site.}
\label{fig:1}
\end{center}
\end{figure}

The details of the VLF/LF transmitting and receiving sites and the path geometries are provided in Table \ref{tab:1}. Locations of the transmitters and receiving site are presented in Figure \ref{fig:1}.

\begin{table*}
\begin{center}
\small
\caption{VLF/LF Transmitting and Receiving Sites}
\begin{tabular}{|c|c|c|c|c|c|c|}
\hline
 & \begin{tabular}[c]{@{}l@{}}Freq\\ {[}kHz{]}\end{tabular}& Country &\begin{tabular}[c]{@{}l@{}}Geographic \\ Latitude {[}deg{]}\end{tabular} &\begin{tabular}[c]{@{}l@{}}Geographic\\  Longitude {[}deg{]}\end{tabular} & \begin{tabular}[c]{@{}l@{}}GCP {[}km{]}\end{tabular} & \begin{tabular}[c]{@{}l@{}}Orientation of\\  propagation path\end{tabular} \\ \hline \hline
Transmitter: GQD &  22.10 & UK & 54.73 N & 2.88 W & 1982 & northwest to southeast \\ \hline
Transmitter: DHO & 23.40 & Germany & 53.08 N & 7.61 E & 1300 & northwest to southeast \\ \hline
Transmitter: ICV & 20.27 & Italy & 40.92 N & 9.73 E & 976 & southwest to northeast \\ \hline
Transmitter: NSC &45.90 & Italy & 38.00 N & 13.50 E & 953 & southwest to northeast \\ \hline
Receiver: AWESOME &  & Serbia & 44.85 N & 20.38 E &  &  \\ \hline
\end{tabular}
\label{tab:1}
\end{center}
\end{table*}

The analysis of VLF/LF data was done together with the examination of the corresponding solar X-ray fluxes. This work deals with a typical X-ray irradiance   $I_{X}$ in $\textrm{Wm}^{-2}$ recorded by GOES-15 satellite in the band 0.1-0.8 nm, available from National Oceanic and Atmospheric Administration USA, via the web site: www.swpc.noaa.gov\-/ftpmenu/lists/xray.html

\subsection{Background ionospheric condition} The accuracy in the description of the ionospheric medium is crucial and the electron density profile, as an important part of its description, is worth of our attention.  The background daytime exponential profile of electron density in general use for VLF modeling \citep{wai64} is given by:

\begin{equation}
\label{eq:Ne}
N_{e}(h, \beta, H')=1.43\cdot10^{13} e^{(0.15H')} e^{[(\beta-0.15)\cdot(h-H')]} \quad \text{m}^{-3},
\end{equation}
with $H'$ in km and $\beta$ in km$^{-1}$. This equation has been successfully used in VLF measurements \citep{tho93,mcr00,mcr04,tho01}. The D-region electron density profile is characterized by the two Wait's parameters: $H'$, as a measure of the reflection height and $\beta$ as a measure of the sharpness or rate of changes of electron density with height. We also use equation (\ref{eq:Ne}) in our work to calculate the altitude density profile in the range 50-90 km.

\subsection{Method of simulations VLF/LF radio signals propagation} The \underbar{L}ong \underbar{W}ave \underbar{P}ropagation \underbar{C}apability waveguide code, LWPC program package \citep{fer90} is used for simulation of VLF/LF propagation along any particular great circle path under different diurnal, seasonal and solar cycle variations in the ionosphere. The LWPC program typically performs the calculations for ten or more modes and has been tested against experimental data. Also, the LWPC program can take arbitrary electron density versus altitude profiles supplied by the user to describe the D-region profile and thus the ceiling of the waveguide.

Using the LWPC code the propagation path of VLF/LF signal was simulated in normal ionospheric condition, with goal to estimate the best fitting pairs of Wait's parameters $\beta_{nor}$ and $H'_{nor}$ (where, \textit{nor} means normal condition) to obtain values of amplitude and phase closest to the measured data for selected day \citep{tho93,tho00,mcr00,zig07}. The next step was to simulate propagation of VLF/LF radio signal through the waveguide in the perturbed D-region induced by additional X-ray radiation for selected moments during the flare duration. In our study we have accepted the presented method and used the observed VLF/LF data to examine the amplitude and phase perturbations during the solar X-ray flare. We used the RANGE model of the LWPC code for examination the single propagation path and specified a range-dependent ionospheric variation. Electron densities were determined from the observed amplitude and phase perturbations by a trial and error method in which electron density profile was modified until the calculated amplitude and phase perturbations matched with observed data. In this manner, the obtained Wait's parameters $\beta_{per}$ and $H'_{per}$ (\textit{per} means perturbed condition) were used for our further calculations.

\section{Investigations on diurnal and seasonal amplitude variations on VLF/LF radio signals}

In literature, the first results about diurnal amplitude variations on VLF signals were published in 1933. \citet{yok33} studied propagation of 17.7 kHz and 22.9 kHz, over long distances $D >5$Mm. They gave explanation for amplitude fading based on single-ray geometrics optics. \citet{bud61} and \citet{wai62} suggested that many rays are needed to explain VLF propagation over long paths. \citet{cro64} and \citet{wal65} put forward an explanation based on the use of modes in the Earth-ionosphere waveguide in which two modes are presented in the nighttime part of the path and only one mode in daylight. Later \citet{cli99} presented studies of VLF propagation over long path,  NAA/24 kHz radio propagating  from Cutler Maine, USA to Faraday, Antarctica during period 1990-1995. VLF radio signal propagated from North to South. They found the times of the amplitude minima were consistent with modal conversation taking place as the day-night  boundary crossed the propagation path at specific locations.

\citet{vol64} presented the studies of diurnal phase and amplitude variation of VLF radio signal at medium distances where the propagation of VLF radio signal did not take place predominantly by one mode. The results were obtained on measurements of VLF data over propagation paths with distances in the range between 300 km and 3000 km at daytime medium,  and at nighttime medium this range was larger. The focus was restricted to regular changes including sunrise effects and solar flare effects. At medium distances the sunrise effects were  very regular and marked in phase and amplitude of VLF radio signals. The sunset effects were much weaker and not so regular.

\subsection{Diurnal amplitude variations}  
\begin{figure*}
\begin{center}
\includegraphics[width=0.4\textwidth]{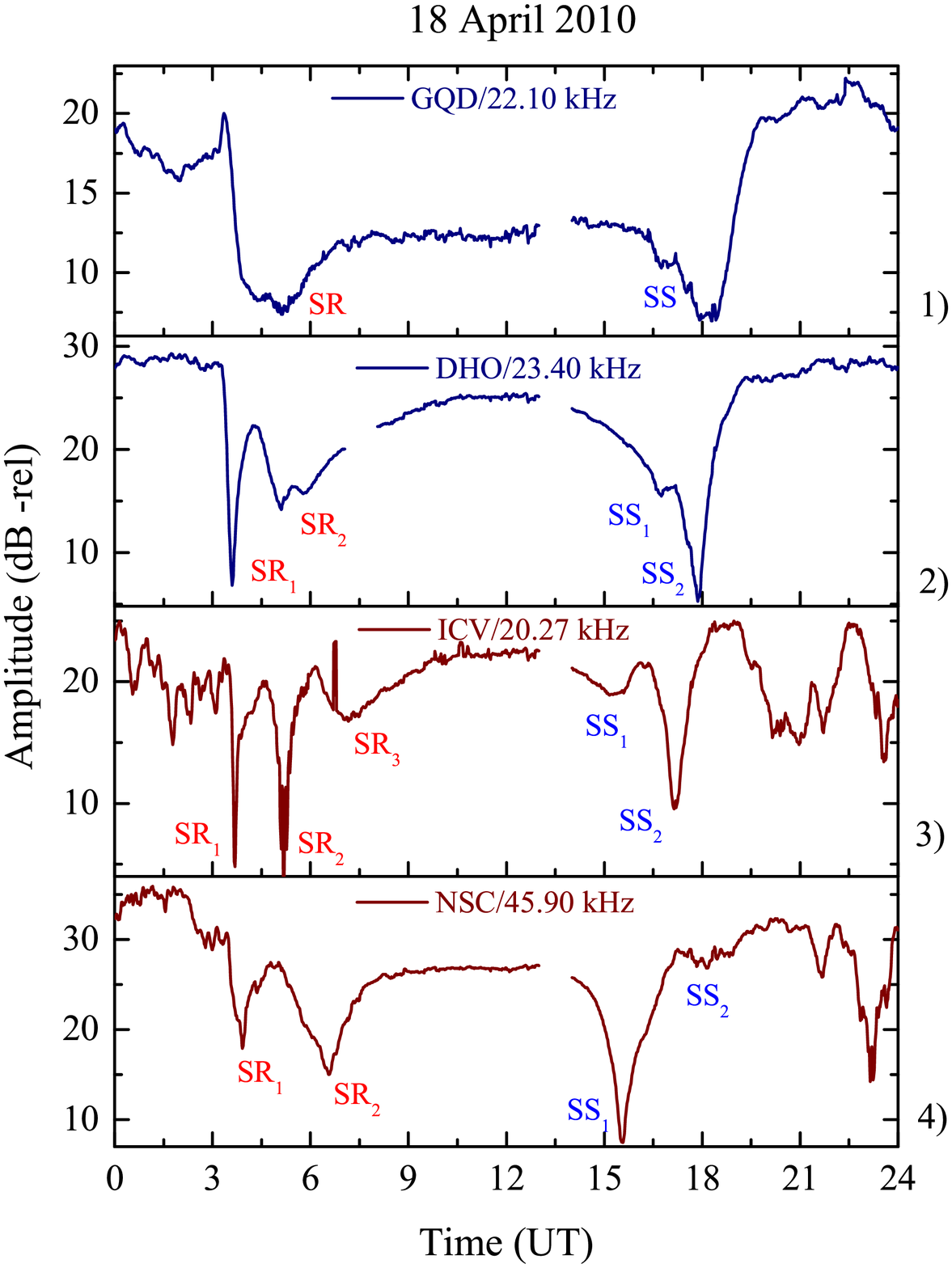}
\includegraphics[width=0.4\textwidth]{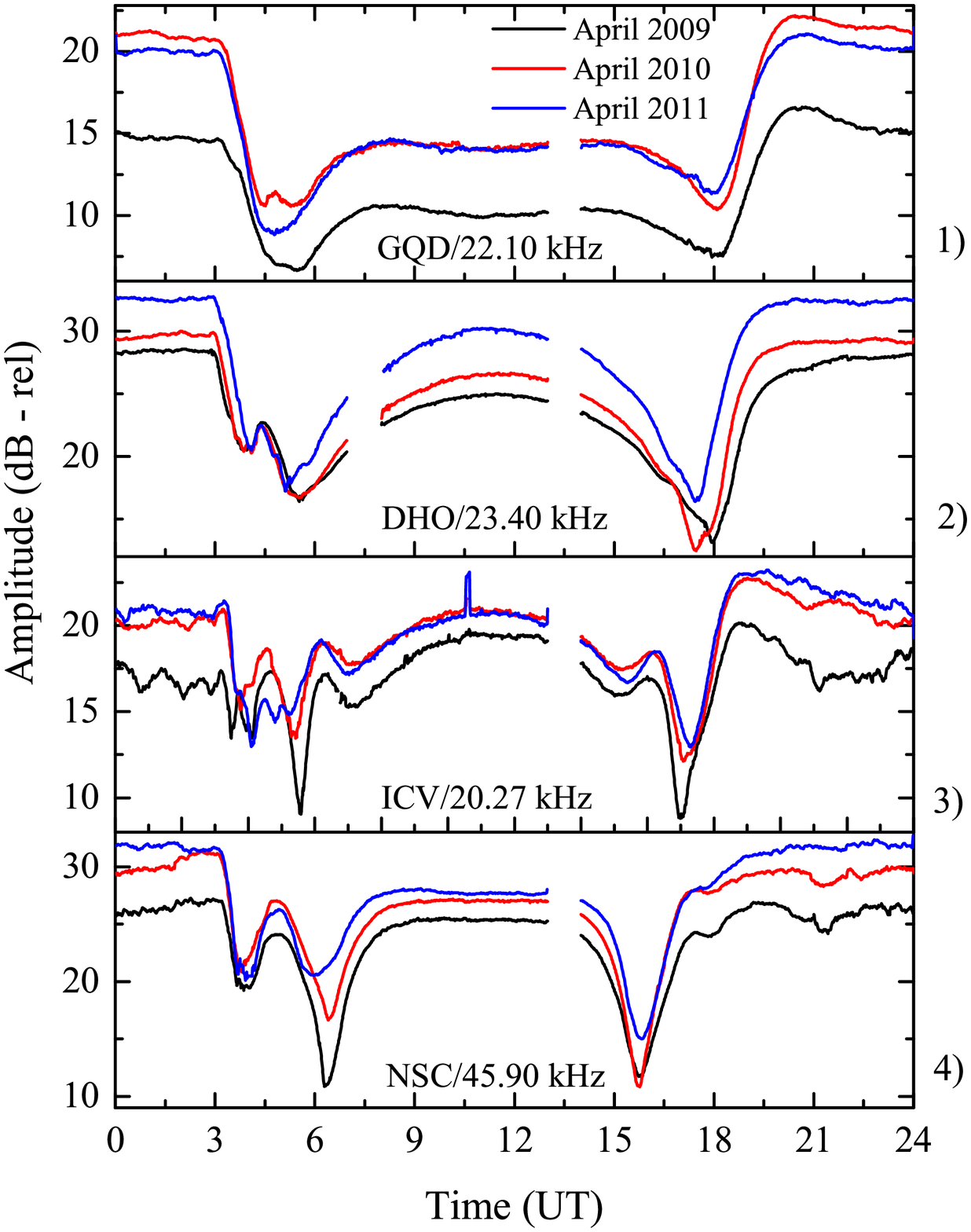}
\caption{\hspace{2in} a) \hspace{2in} b) \newline a) Diurnal variation of amplitude on VLF/LF signals at GQD/22.10 kHz, DHO/23.40 kHz, ICV/20.27 kHz and NSC/45.90 kHz monitored on 18 April 2010, b) Variations of monthly averaged amplitude of VLF/LF signals obtained for April 2009, 2010 and 2011.}
\label{fig:2}
\end{center}
\end{figure*}

Diurnal behaviors have been examined at different frequencies all monitored at Belgrade site. For this purpose day 18 April 2010 was selected as representative day of normal ionospheric conditions under low solar activity. The Earth-facing side of the Sun was blank, without sunspots. During that day at 02:05 UT the maximum of X-ray irradiance $I_{Xmax} = 10^{-7}$ Wm$^{-2}$ in the band 0.1-0.8 nm, was recorded. Diurnal variation of amplitude on GQD/22.10 kHz, DHO/34.40 kHz, ICV/22.27 kHz and NSC/45.90 kHz radio signals over 24 hours are shown in Figure \ref{fig:2}a. Characteristic events of amplitude minima that occurred during the sunrise and sunset were identified for each path and marked with SR and SS on Figure \ref{fig:2}a. There is a data gap between 13:00-14:00 UT on each trace caused by data archival. Also, there is data gap between 07:00-08:00 UT on amplitude variations of DHO/23.40 kHz radio signal, because in that period the transmitter is off-air. The shape of the expected signal curve varies dramatically over the path with time.

For comparison, we calculated monthly averaged amplitude values of GQD/22.10 kHz, DHO/23.40 kHz, ICV/20.27 kHz and NSC/45.90 kHz radio signals for April 2009, 2010 and 2011. Figure \ref{fig:2}b shows the monthly averaged amplitude values for four VLF/LF radio signals against universal time over 24 hours. In the averaged data for ICV there are spikes at $\sim$ 10:40  UT in both the April 2010 and April 2011 curves. Sporadically amplitude of ICV/20.27 kHz increases by $\sim$ 2 dB in duration of few minutes. As this appearance is repeatable always at $\sim$ 10:40 UT, the source of it is artificial (Figure \ref{fig:2}b, panel number 3).  In April 2011 the solar X-ray flares occurred during a few days, and these days were eliminated from calculations. To present development of solar activity during the current solar cycle 24 we cite \underline{s}moothed \underline{s}unspot \underline{n}umbers (SSN) for April 2009, 2010 and 2011 and they are: 2.2, 15.4 and 41.8, respectively (http://www.ips.gov.au/Solar/1/6).

GQD/22.10 kHz radio signal propagates from Skelton UK to Belgrade and this path is far apart in longitude in comparing with other paths analyzed in this paper. This is West-East propagation and the distance between transmitter and receiving site is $D =$ 1982 km. The first panel of Figure \ref{fig:2}a shows the observed diurnal variation of the amplitude on VLF radio signal at 22.10 kHz. There is well evidence between amplitude recorded during nighttime and daytime conditions. The variation of amplitude has larger values during nighttime than in daytime condition, because of lower absorption in the D-region. During the transition between nighttime/daytime propagation conditions, two not sharp minima in amplitude, labeled by SR and SS are seen in Figure \ref{fig:2}a panel number 1.

The first panel of Figure \ref{fig:2}b shows monthly averaged amplitude values of GQD/22.10 kHz radio signal over 24 hours by order of succession three years. The shapes of the curves are very similar to each other. On each curve, there are well defined two amplitude minima. The time of development the amplitude minima are repeatable from year to year. Monthly averaged amplitude values for April 2010 and 2011 are greater in comparing with values for April 2009, and we assume it was induced by the changes in solar activity.

\begin{figure*}
\begin{center}
\includegraphics[width=0.43\textwidth]{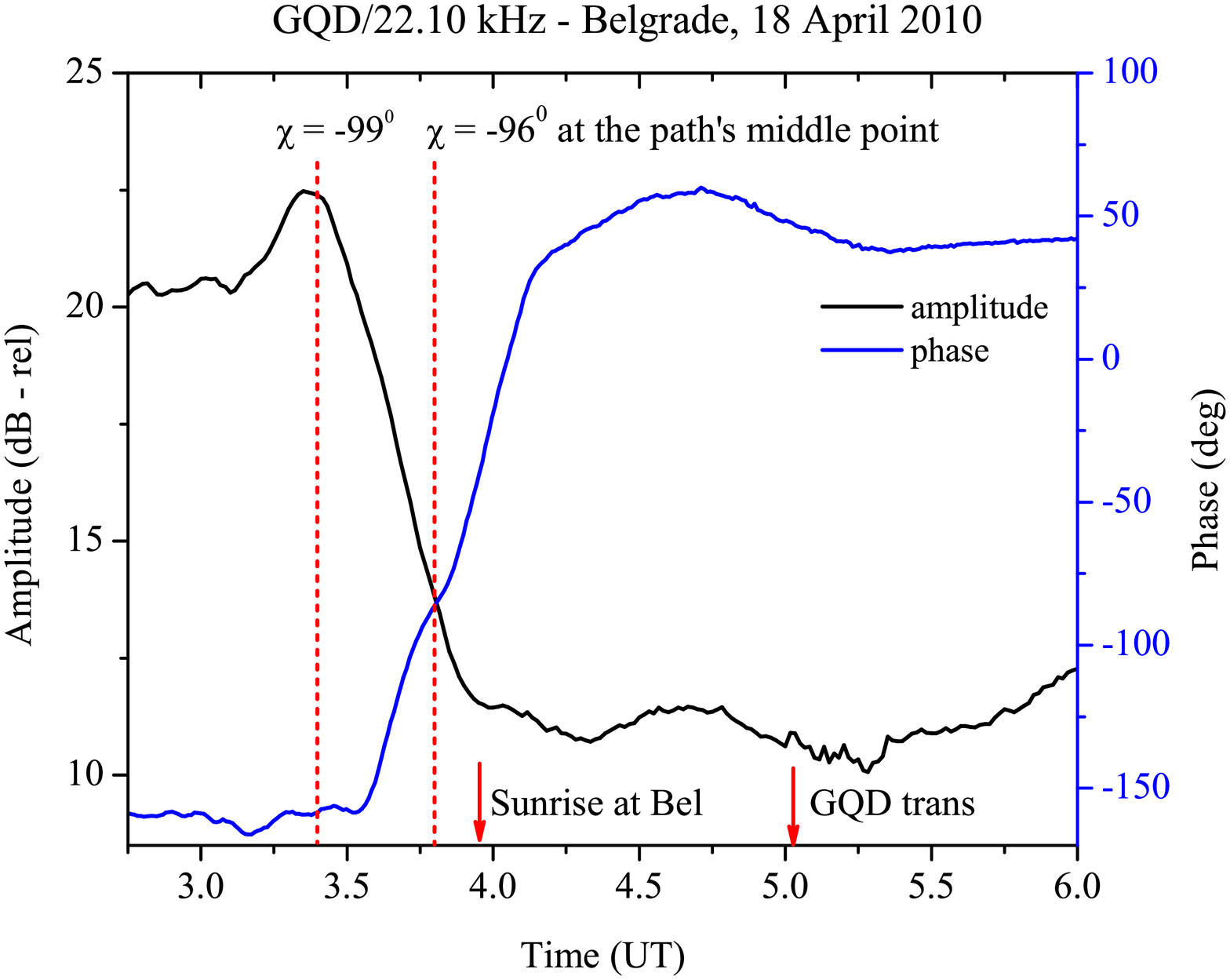}
\includegraphics[width=0.43\textwidth]{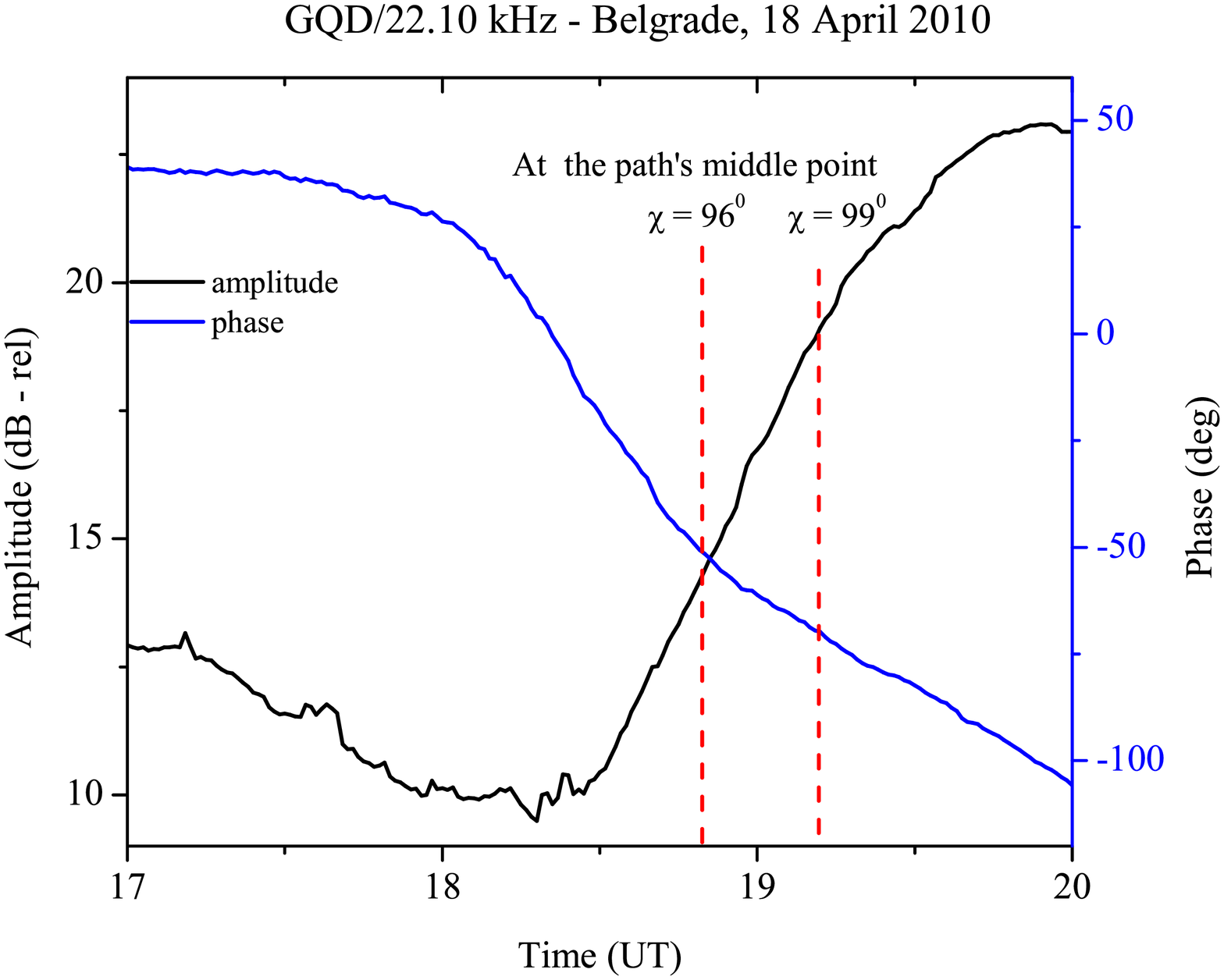}
\caption{\hspace{2in} a) \hspace{2in} b) \newline a) Amplitude and phase on GQD/22.10 kHz during sunrise recorded at Belgrade on 18 April 2010 b) Amplitude and phase on GQD/22.10 kHz during sunset at Belgrade on the same day. For convenience, morning solar zenith angles are shown as negative while those for the afternoon are positive.}
\label{fig:3}
\end{center}
\end{figure*}

\citet{gru08} presented measured and calculated diurnal variations of amplitude and phase on GQD/22.10 kHz radio signal recorded at Belgrade site for one summer day in 2005. Considering the results, it is evident that the diurnal phase variation of GQD/22.10 kHz radio signal is generally in the form of a trapezium, where all night conditions over the path correspond to one steady-phase level and all day conditions over the path correspond to the other steady-phase level. In all our examinations of diurnal phase variations on GQD/22.10 kHz recorded at Belgrade site from 2008 to 2014 we obtained a similar shape of curves with the example presented in the paper \citep{gru08}. The idealized transition from one phase level to the other during sunrise and sunset completes the straight sides of the trapezium.  In this paper we will present the phase steps during transition periods for recorded data on 18 April 2010. Observed variations of amplitude and phase on GQD/22.10 kHz radio signal over three hours are presented on Figure \ref{fig:3}a and \ref{fig:3}b.

We  accepted  the results obtained by \citet{vol64} that VLF radio signals  propagating  from transmitter to receiver over paths with distance $\sim$ 2000 km reflect once on the middle of the paths. By the LWPC code we simulated the propagation of GQD/22.10 kHz radio signal and obtained coordinates 50.30$^{0}$ N, 10$^{0}$ E for the middle of the propagation path, and number of discrete modes under different diurnal condition over path.
For convenience, morning solar zenith angles are shown as negative while those for the afternoon are positive.
The process of ionization in the D-region begins when solar zenith angle has value $\chi = -99^{0}$, and sunrise terminator reaches a height $h$ = 95 km. The next important moment is when sunrise terminator reaches a height  $h$ = 35 km that occurs for $\chi = -96^{0}$.  In time interval that corresponds to values of the solar zenith angle $\chi = -99^{0}$ and $\chi = -96^{0}$ there are changes of the altitude profile of ionospheric conductivity against time. By changing the time as input parameter with the step of one minute we defined the times of solar zenith angles $\chi = -99^{0}$ and $\chi = -96^{0}$ at the middle of the propagation path.  In this way we got the information about the time of sunrise at $h$ = 95 km and $h$ = 35 km in the D-region. Changes in the conductivity of the D-region cause variations in amplitude and phase of the VLF/LF radio signals propagating across the terminator line. These moments are marked with red dashed lines on Figure \ref{fig:3}a. Also, the sunrises at Belgrade site and GQD transmitter (on ground level) are marked with red arrows. VLF radio signal propagates from nighttime to daytime conditions, with number of discrete modes $n_{n}$ =17 and $n_{d}$ =7, respectively.
Figure \ref{fig:3}a shows the transition from phase level during nighttime to phase level during daytime, starting after the sunrise occurs at $h$ = 95 km. Simultaneously with phase step the development of the amplitude minimum is presented on Figure \ref{fig:3}a.

\begin{figure*}
\begin{center}
\includegraphics[width=\columnwidth, height=0.75\columnwidth]{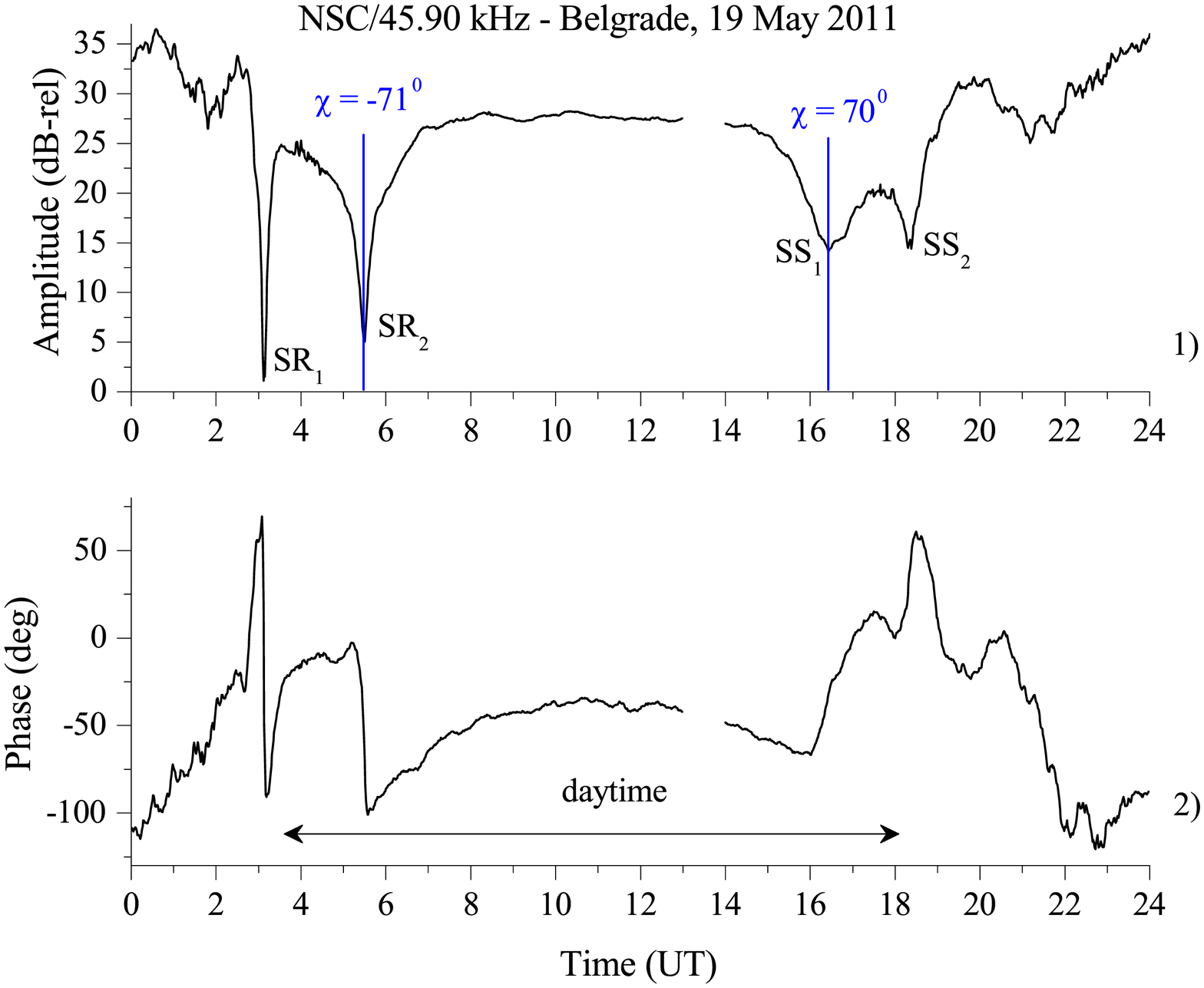}
\caption{Diurnal variations of amplitude and phase on NSC/45.90 radio signal against universal time recorded at Belgrade site during 19 May 2011.}
\label{fig:4}
\end{center}
\end{figure*}

In confirmation of our assumption that GQD/22.10 kHz radio signal once reflects from the ionosphere (one-hop) along path, $D$ = 1982 km is in correlation timing of the creating the first minimum with time interval of the illumination ionosphere in the altitude range, 95-35 km  in the middle of the path.

During sunset the opposite changes occur in the D-region. Figure \ref{fig:3}b shows the transition from phase level during daytime to phase level during nighttime and the development of amplitude minimum. We defined the times when sunset terminators reaches height at $h$ = 35 km and  $h$ = 95 km  at the middle of the propagation path. Red dashed lines indicate these times on Figure \ref{fig:3}b. The amplitude minimum occurred at $\sim$ 18:15 UT about one hour earlier than sunset is at $h$ = 95 km.

DHO/23.40 kHz radio signal propagates from Rhauderhent, Germany to the Belgrade site across an all land path. VLF radio signal propagates Northwest-Southeast and the distance between transmitter and receiver site is $D =$ 1300 km. The second panel of Figure \ref{fig:2}a shows the observed variations of amplitude on DHO/23.40 kHz radio signal against time over 24 hours. The amplitude of VLF radio signal varies in a characteristic way that is defined by geophysical parameters of transmitter and receiver site. The differences in amplitude values recorded during nighttime and daytime conditions are evident.
Four amplitude minima labeled as SR$_{1}$, SR$_{2}$, SS$_{1}$ and SS$_{2}$ are observed, respectively during sunrise and sunset transition  along the propagation path.
The amplitude of the signal is generally dependent on a superposition of discrete modes (nighttime: $n_{n}$ = 18 and daytime: $n_{d}$ =7), which depends on the variations of the waveguide parameters. The amplitude minima are produced by modal interference generated at the sunrise and sunset height discontinuities in reflection height as they move along the path \citep{wal65}.

At the middle of the path (49$^{0}$N, 14.5$^{0}$E) sunrise reaches height $h$ = 95 km at 03:13 UT and $h$ = 35 km at 03:33 UT on 18 April 2010. From recorded data it is evident that amplitude started to fall from nighttime level at $\sim$ 03:15 UT and had minimum value at 03:36 UT. Development of amplitude minimum SR$_{1}$ is in good correlation with changes of illumination at the middle of the path. The amplitude had minimum value SS$_{2}$ at 17:55 UT. The sunset reaches height $h$ = 95 km at 18:46 UT and than amplitude value is very close to values of nighttime level. During daytime condition over DHO-BEL path there are two amplitude minima SR$_{2}$ (morning) and SS$_{1}$ (afternoon) developed under solar zenith angles $\chi = -81^{0}$ and $\chi = 80^{0}$, respectively.

The second panel of Figure \ref{fig:2}b shows monthly averaged values of amplitude on DHO/23.40 kHz for April 2009, 2010 and 2011. The shapes of curves which presented monthly average variations of amplitude over 24 hours are very similar. There are some differences in values from year to year. Also four amplitude minima are noticeable.

Radio signals with frequency ICV/20.27 kHz and NSC/ 45.90 kHz propagete from Southwest to Northeast over short paths 976 km and 953 km, respectively. Both radio signals propagate over sea, land, sea and land, which implies to very similar conductivity properties of the waveguide bottoms. The third panel of Figure \ref{fig:2}a shows amplitude variation on ICV/20.27 kHz  radio signal against universal time. For this VLF radio signal the differences in amplitude values recorded during nighttime and daytime conditions are not well recognized (nighttime: $n_{n}$ = 16 and daytime: $n_{d}$ =7).
Three amplitude minima labeled as SR$_{1}$ SR$_{2}$ and SR$_{3}$ are observed during sunrise transition and morning over the propagation path. Two of them are very sharp. Two minima amplitude labeled by SS$_{1}$ and SS$_{2}$ are observed during afternoon and sunset transition, respectively. The first amplitude minimum SR$_{1}$ is very sharp and occurs in time interval close to time interval during which sunrise terminator reaches middle of the propagation path (42.96$^{0}$ N, 14.77$^{0}$ E).

For studying diurnal variations on ICV/20.27 kHz radio signal against time during three year period we calculated monthly averaged amplitude values and the results are shown on the third panel of Figure \ref{fig:2}b. All the curves for April 2009, 2010 and 2011 have similar shape with well defined five amplitude minima on each of them.

The fourth panel of Figure \ref{fig:2}a shows diurnal variations of amplitude on NSC/45.90 kHz radio signal against time recorded on 18 April 2010. The amplitude has larger values during nighttime than in daytime condition, due to lower absorption during night. Also four amplitude minima were developed during that day. The fourth panel of Figure \ref{fig:2}b shows monthly averaged amplitude data against time. In all the curves it is possible easily to define the amplitude minima. Also, the times of their development are repeatable from year to year.

\begin{figure*}
\begin{center}
\includegraphics[width=0.4\textwidth]{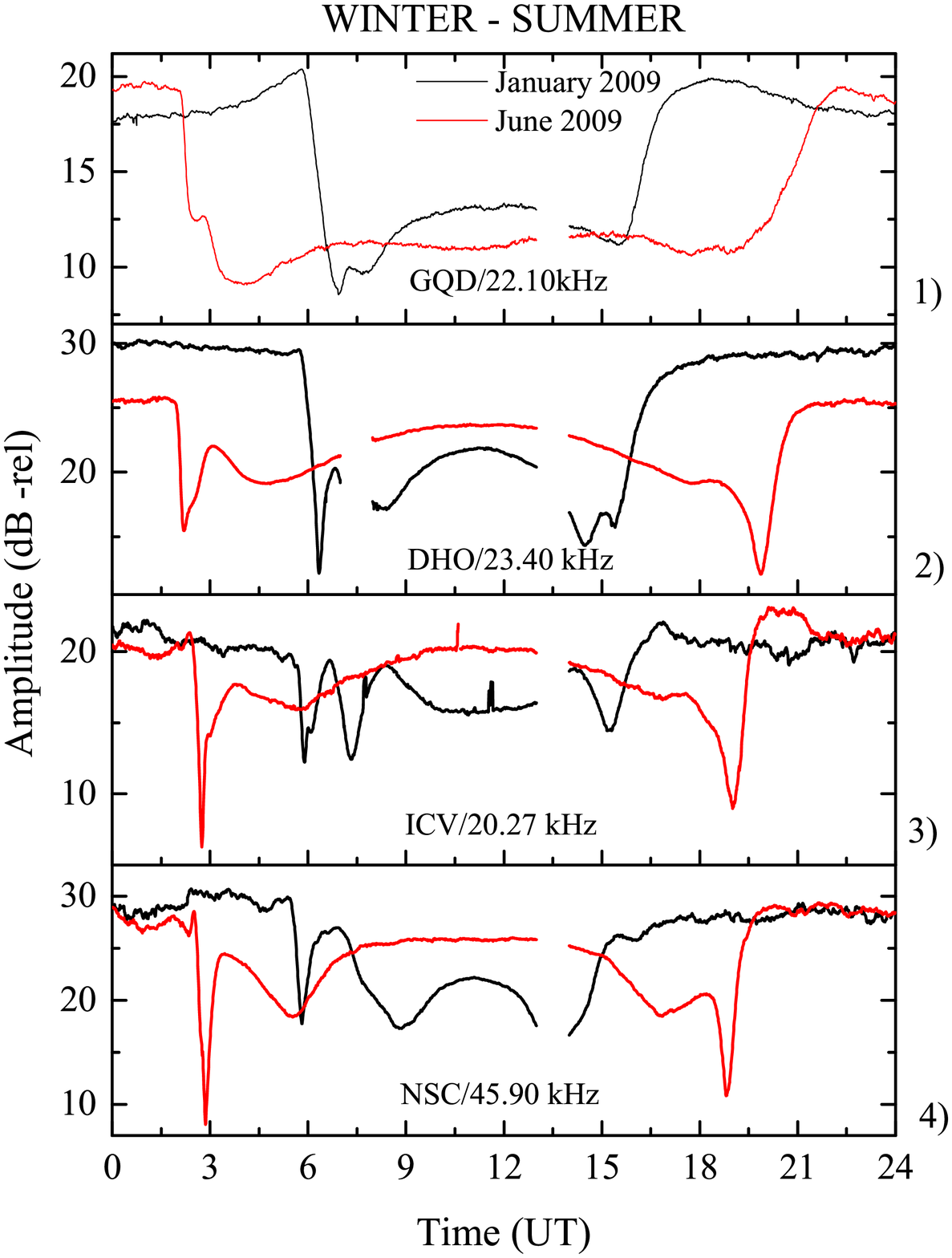}
\includegraphics[width=0.4\textwidth]{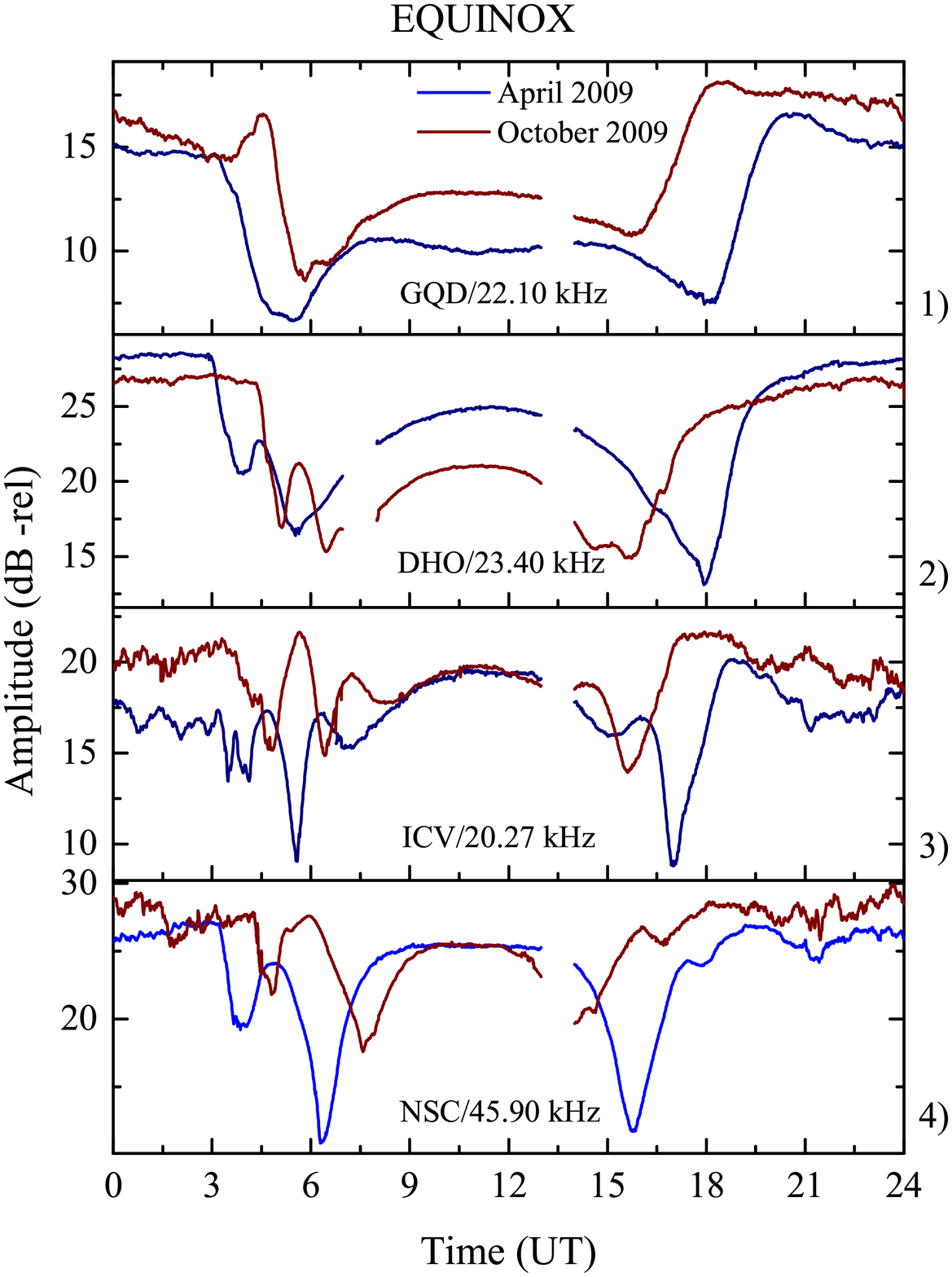}
\caption{\hspace{2in} a) \hspace{2in} b) \newline The monthly averaged amplitude on GQD/22.10 kHz, DHO/23.40 kHz, ICV/20.27 kHz and NSC/45.90 kHz radio signals versus time over 24 hours a) for January and June 2009, b) for April and October, 2009.}
\label{fig:5}
\end{center}
\end{figure*}
Along with the study of diurnal variation of amplitude we have examined the diurnal variation of phase on NSC/45.90 kHz radio signal. This LF radio signal propagates from Southwest to Northeast over short path, $D= 953$ km. By the LWPC code we simulated the propagation of NSC/45.90 kHz radio signal and obtained times of sunrise/sunset at heights: $h = 95$ km and $h = 35$ km in the D-region for middle part of the propagation path (41.53$^{0}$ N, 16.75$^{0}$ E). Our numerical results show that LF radio signal propagates from nighttime to daytime condition as a superposition of discrete modes $n_{n} $ $\le$ 34 and $n_{d}$ = 10, respectively. Sunrise effects on LF propagating cause a gradual fall of number of discrete modes over a short path.
Data of amplitude and phase on NSC/45.90 kHz radio signal recorded on 19 May 2011 are given on Figure \ref{fig:4}. The first panel of Figure \ref{fig:4}  shows amplitude variation on LF radio signal against universal time with well developed four amplitude minima. The second panel of Figure \ref{fig:4} presents monitored phase data for the same day. During dawn and dusk time sector there are not smoothly transitions of phase from one stable level to another.

Timing of development the first amplitude minimum, SR$_{1}$ is in good correlation with time when sunrise terminator reaches the middle part of the propagation path. The phase transition from nighttime to daytime condition passes through the peak, while the amplitude drops to lower values. In the process of changing daytime to nighttime condition, when sunset terminator reaches middle part of path there is occurrence of the last amplitude minimum, SS$_{2}$. Development of the amplitude minimum SS$_{2}$ is followed with well recognized peak in phase values.

During daytime condition over path, two amplitude minima labeled as SR$_{2}$  (morning) and SS$_{1}$  (afternoon) are observed.  The lowest values of these amplitude minima are recorded under solar zenith angle $\chi = -74^{0}$ and $\chi = 71^{0}$, respectively. The amplitude minimum SR$_{2}$   is followed with suddenly decreasing of phase values, while the amplitude minimum SS$_{1}$   is in correlation with increasing phase values.

DHO-BEL, ICV-BEL and NSC-BEL paths and great segment of GQD-BEL path are in the same time zone. Middle of GQD-BEL and NSC-BEL propagation paths are far apart in longitude for $7^{0}$.
All the paths are similarly illuminated during daytime condition while there are differences in the level of illumination during dawn and dusk in accordance to geographic coordinates of transmitter.
Based on these facts, our results are as follows:

\begin{enumerate}


\item The process of ionization in the D-region begins when solar zenith angle reaches value $ \chi = -99^{0}$, and sunrise terminator reaches the height $h = 95$ km. When this process starts in the middle of the path, it causes the changes of altitude profile of ionospheric conductivity and the appearance of the first amplitude minimum.

\item Our results based on the simulation of the VLF/LF propagation over short paths ($D$ $<$ 2 Mm) show that 20.27 kHz, 22.10 kHz and 23.40 kHz radio signals propagate under nighttime condition as a superposition of 16, 17 and 18 discrete modes, respectively. Under daytime condition all of these VLF radio signals propagate as a superposition of 7 discrete modes.
    Our numerical results show that LF radio signal propagates from nighttime to daytime condition as a superposition of discrete modes $n_{n}$ $\sim$ 34 and $n_{d}$ = 10, respectively. Sunrise effects on LF propagating over a short path cause a gradual fall of number of discrete modes.

\item How many amplitude minima and at what time they would be developed are based on specifications for each path. The occurrences of amplitude minima depend on the interferences of various number of discreet modes. All possible combinations of conditions: \textit{sunrise and sunset, position of terminator on propagation path, relative positions of transmitter and receiver, distance, ground conductivity and transmitted frequency} are responsible for the occurrence of amplitude minima.

\item The main point of our result is that the amplitude minima could be divided in two types: The amplitude minima that are developed in time intervals during transition of nighttime/daytime and daytime/nighttime conditions on the middle of the propagation path belong to the first type. The second type of amplitude minima occurs under daytime condition over all short paths.
    Amplitude minima of second type usually appear as a pair, one during morning and other during afternoon and their timings are symmetrical arrange in a according to a local noon. Timings of their occurrences continuously change from day to day.
\end{enumerate}


\subsection{Seasonal amplitude variations}  We examined four VLF/LF radio signals recorded at Belgrade site to follow main propagation characteristics induced by different levels of illumination over 24 hours and over one year. For this purpose we analyzed the measurements obtained in the years of low solar activity. Figure \ref{fig:5}a shows the averaged amplitude variations against time for winter month, January  and summer month, June 2009. There is a gradual shift between winter and summer amplitude levels on radio signals: DHO/34.40 kHz, ICV/20.27 kHz and NSC/45.90 kHz. Monthly averaged amplitudes on GQD/22.10 kHz have larger values during winter than summer months. All VLF propagation paths are differently illuminated under winter and summer conditions causing time differences in the appearances of amplitude minima.

Figure \ref{fig:5}b shows monthly averaged amplitude on VLF/LF radio signals for equinox months: April and October 2009. As for these two months, on the basis of comparing the curves for monthly average amplitude values during the period of 24 hours, it is obvious that the shapes of these curves are similar to each other. There is a time difference when it comes to the moment of appearance and development of amplitude minima in April and October 2009.

On the basis of measured VLF/LF data our results for short paths, $D < 2$ km are:

\begin{itemize}
  \item[] Timings of development the amplitude minima when sunrise and sunset terminators reach middle of the propagation path are in correlation with seasonal effects on the D-region. During daytime there are one or two pairs of amplitude minima whose creation is a consequence of destructive interference of modes. The temporal variability in their creation is correlated with number of light hours for each day during year. The amplitude minima in dawn and morning are sharper than in afternoon and dusk and are the sharpest during summer.
\end{itemize}


\section{Typical amplitude and phase changes of\\ VLF/LF radio signals induced by solar flares }

A characteristic of the radiation emission in the EUV, X-ray and radio signals is that it is not uniformly distributed in the Sun's corona, but mostly concentrated in localized emission centers. These emission centers belong to a class of phenomena known collectively as \textit{active regions}. A solar flare is a sudden, rapid and intensive phenomenon in solar activity, releasing a large amount of energy (up to about 10$^{25}$J) in the solar atmosphere \citep{pro04,bot07}. However, when a solar X-ray flare occurs, there is a major increase in the flux of X-rays from the Sun and those with wavelengths appreciable below 1nm are able to penetrate down to the D-region heights and cause there additional ionization. The rapid increase in the electron density leads to several phenomena grouped together under the name \textit{\underbar{s}udden \underbar{i}onospheric \underbar{d}isturbance }(SID). The increased electron density lowers the effective reflection height $H'$ and causes a very strong influence on propagation characteristics of VLF/LF radio waves \citep{mit74}.

\subsection{Amplitude variations due to solar activity cycle} There are times when the active regions are rare or only weakly defined. Under these conditions the Sun is designated as "quiet".  An "active" Sun is characterized by numerous strong active regions. A seven-year period (2008-2014) includes the minimum of solar cycle 23 in December 2008 and the maximum of solar cycle 24 in April 2014, with 82 SSN. Solar cycle 24 belongs to the category of moderate cycles (http://www.ips.gov.au/Solar/1/6). In Table \ref{tab:2} there are some characteristics of solar cycle 24. Following the development of X-ray flares in solar cycle 24 it shows that no flares were observed in January 2009, while 282 solar X-ray flares of C, M and X class were observed in February 2014 (www.tesis.lebedev.ru).

\begin{figure*}
\begin{center}
\includegraphics[width=0.49\textwidth]{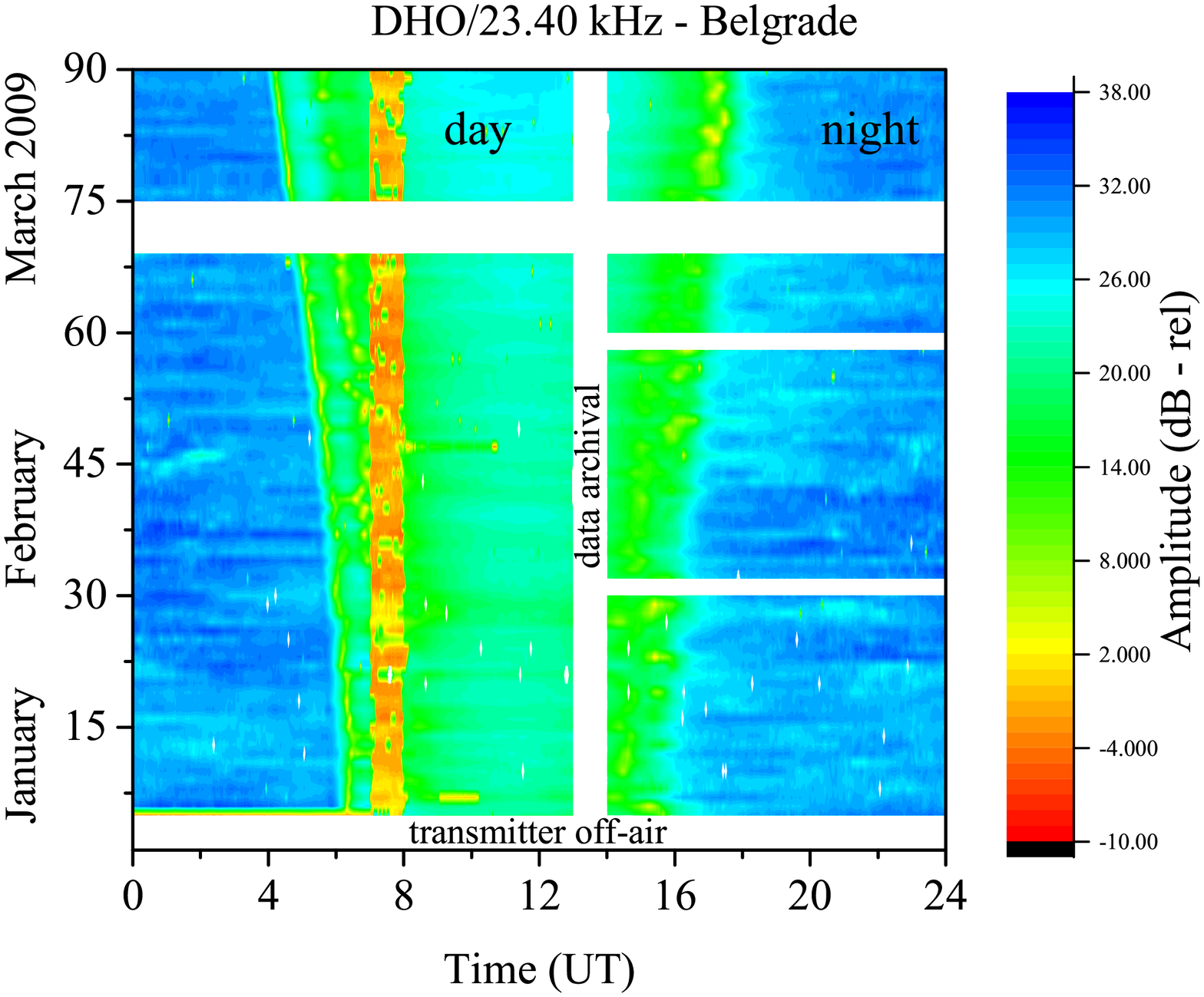}
\includegraphics[width=0.49\textwidth]{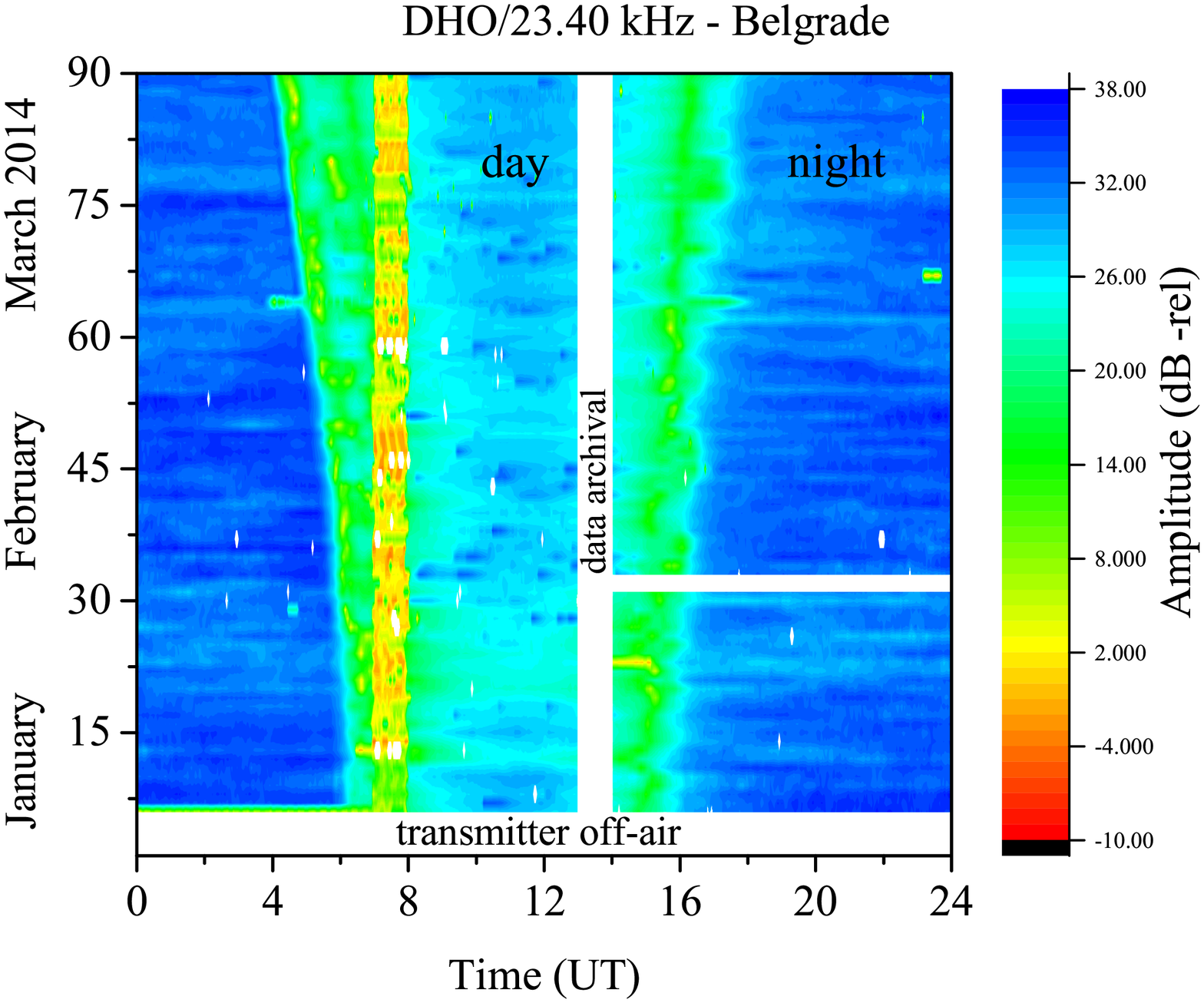}
\caption{\hspace{1in} a) \hspace{3in}   b) \newline  Amplitude  measurements on DHO/23.40 kHz radio signal at Belgrade site during:  a) January-March 2009  b) January-March 2014.}
\label{fig:6}
\end{center}
\end{figure*}

VLF/LF propagation as noted above provides some insight on the way ionization builds up and decays, and provides a routine monitor for the detection and time change of solar-geophysical disturbances. The Figures \ref{fig:6}a and \ref{fig:6}b illustrate the diurnal-seasonal variations of amplitude on DHO/23.40 kHz radio signal in Belgrade for three-month period during "quiet" (2009) and "active" (2014) solar condition. White areas were missing data caused by: transmitter off-air, archival data or failures at the receiver site.

The diurnal-seasonal variations of amplitude are caused by variation of the equivalent reflection height and the reflection characteristics of the D-region during 24 hours. The Figure \ref{fig:6}a shows that amplitudes of signal are typically higher and more variable at night (different shades of blue color) than during the day (equal turquoise color). At dawn and dusk amplitude on DHO/23.40 kHz radio signal passes through the minimum producing sharp border between nighttime and daytime amplitude values (green gradually turns to turquoise color). The dawn crossing (left sides of Figure \ref{fig:6}a and \ref{fig:6}b) is sharper than the dusk crossing. The recorded VLF amplitude can be higher or lower at night, depending on the path, due to the summation of the modes. It is characterized by periodic and repeatable variations of amplitude as the dawn-dusk terminator moves along a DHO-BEL path.

For better interpretation of measured data we simulated VLF propagation under normal ionospheric condition. We selected date: 20 January 2009 and time 06:16 UT to calculate main propagation parameters of DHO/23.40 kHz radio signal from transmitter along path to Belgrade using the LWPC code. The modeling of VLF radio signal propagation in the Earth-ionosphere waveguide in which the illumination smoothly changes is carried out and results are given in Table \ref{tab:3}. Numerical values of: \textit{ground conductivity $\sigma$, solar zenith angle $\chi$, sharpness $\beta$, reflection height $H'$, electron density at $H'$ and number of modes as a function of distance from transmitter along path to receiver} are presented in Table \ref{tab:3}. Analysis of data shows that electron density increases, reflection height moves down in the D-region and the number of discrete modes reduces from $n_{n}$=18 to $n_{d}$=7. The consequence of the occurrence of all these processes in the waveguide is that amplitude of VLF signal during daytime has smaller value than during the night.

The ionospheric effects of X-ray flares provide one major additional source of interest - \textit{the reaction of the ionospheric medium to an impulsive ionization}. Disturbances in the D-region induce increase of amplitude on DHO/23.40 kHz radio signal. Figure \ref{fig:6}b shows recorded amplitudes on DHO/23.40 kHz radio signal against universal time over 24 hours during January-March 2014. Perturbations of amplitudes during daylight hours appear as blue streaks on Figure \ref{fig:6}b. To present changes in the daytime D-region during SIDs we modeled VLF propagation by changing Wait's parameters $\beta$ and $H'$ in RANGE model of LWPC code. The modeling of VLF propagation in the waveguide under normal ionospheric condition and during two SIDs were carried out and results are presented in Table \ref{tab:4}. Recent collections of data provide even more detailed insight into variation of signal amplitude and consequently into ionospheric variability. During a SID, the D-region becomes highly ionized, the altitude profile of ionospheric conductivity changes and VLF radio signal reflects from a lower height and all of these changes result that VLF signal propagates as a superposition of more discrete modes than in normal ionospheric condition. The signal strength of the wave increases and in our modeling amplitude of VLF signal changes from $A_{nor}$ = 31.5 dB to $A_{per}$ = 36.5 dB.

\begin{figure*}
\begin{center}
\includegraphics[width=0.43\textwidth]{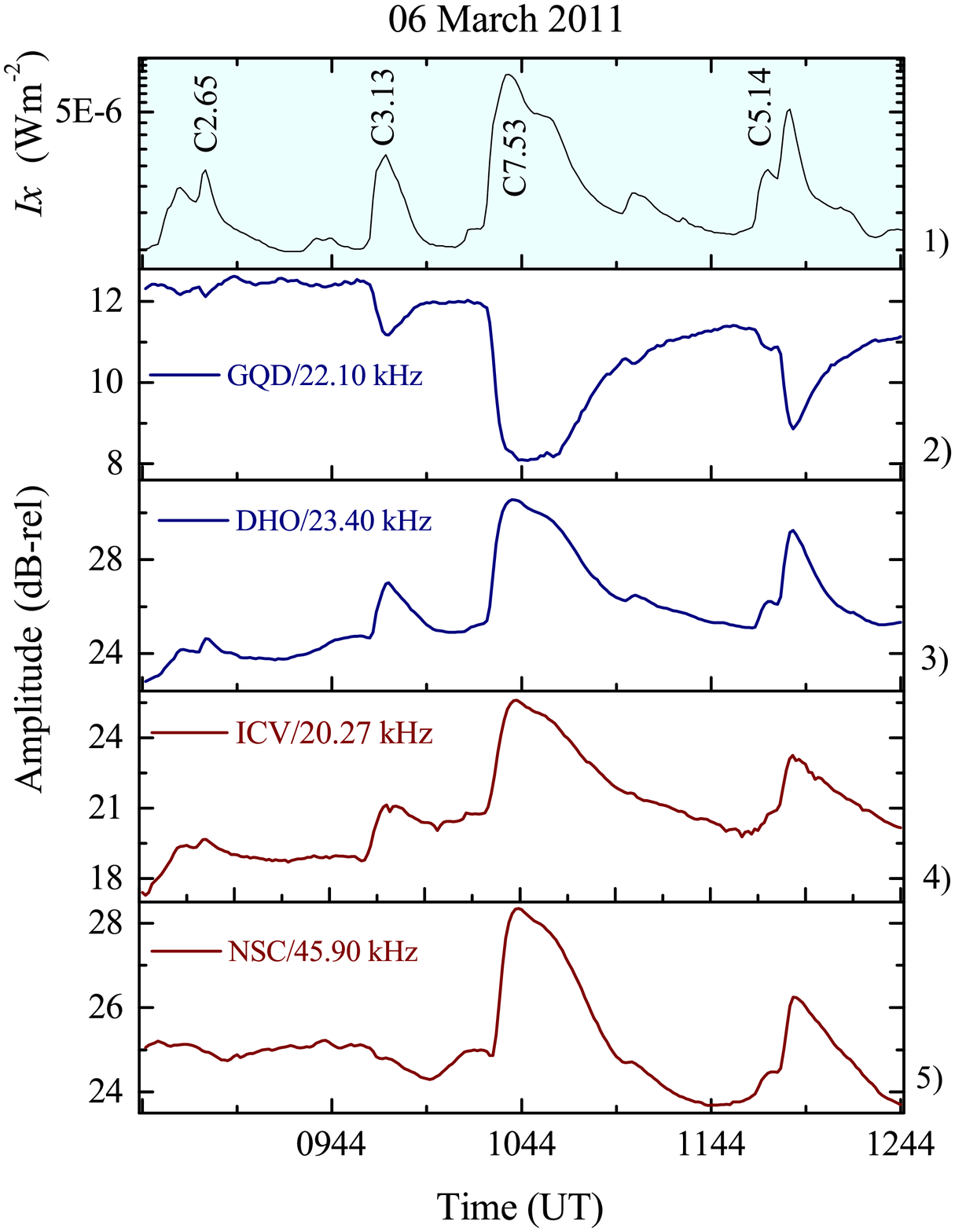}
\includegraphics[width=0.43\textwidth]{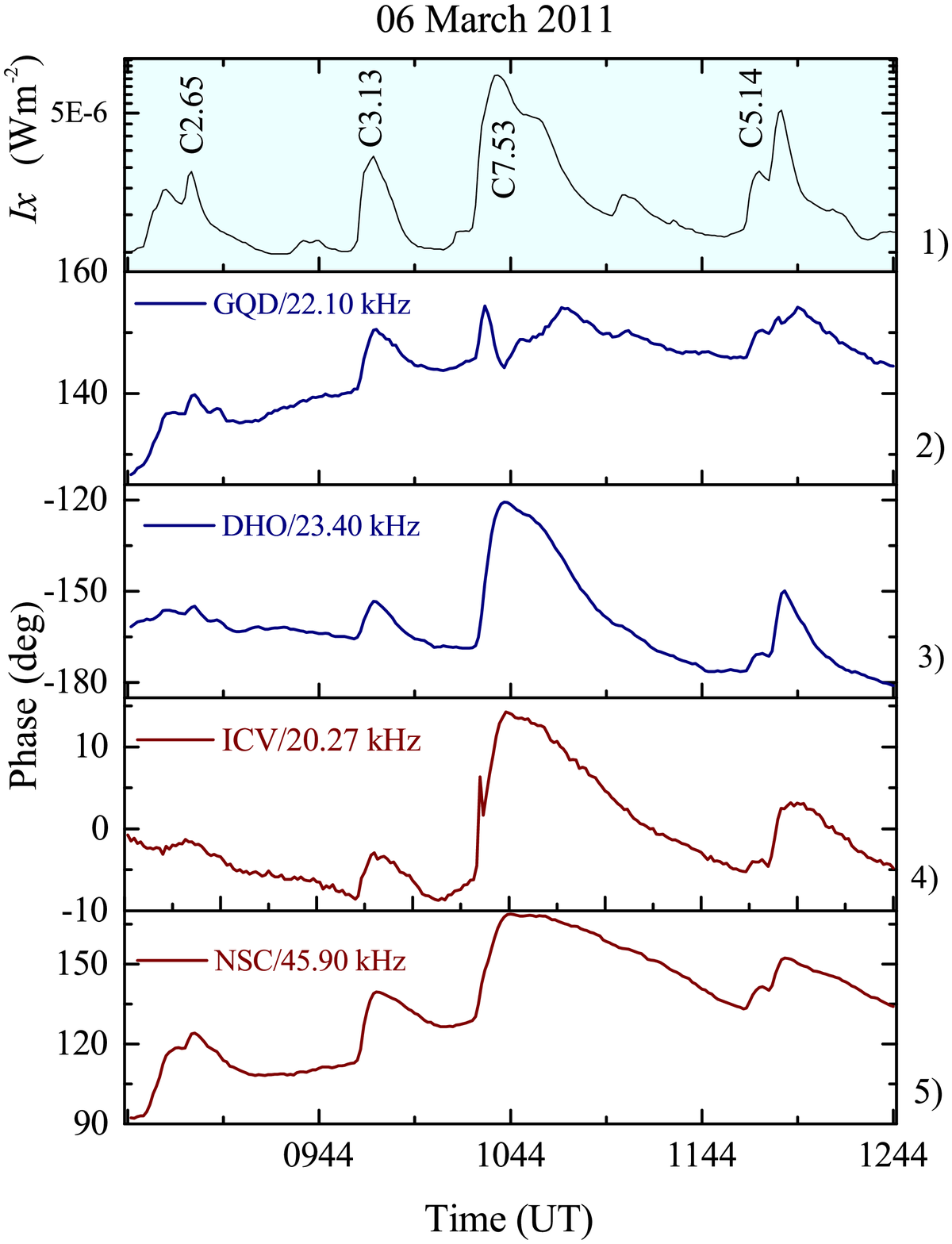}
\caption{\hspace{2in} a) \hspace{2in} b) \newline  Simultaneous variations of X-ray flux, amplitude and phase of GQD/22.10 kHz, DHO/23.40 kHz, ICV/20.27 KHz and NSC/45.90 kHz radio signals against universal time a) perturbation of amplitudes during four successive flares on 6 March 2011 b) perturbation of phases during four successive flares on the same day.}
\label{fig:7}
\end{center}
\end{figure*}

\begin{table}
\small
\centering
\caption{Main characteristics of solar cycle 24}
\label{tab:2}
\begin{tabular}{|c|l|c|l|c|l|l|c|}
\hline
Cycle & \multicolumn{2}{l|}{\begin{tabular}[c]{@{}l@{}}Sol. Start\\ Year   Month\end{tabular}} & \multicolumn{2}{l|}{\begin{tabular}[c]{@{}l@{}}Sol. Max.\\ Year    Month\end{tabular}} & \begin{tabular}[c]{@{}l@{}}Max\\ SSN\end{tabular} & \multicolumn{2}{l|}{\begin{tabular}[c]{@{}l@{}}Rise to Max\\ Years      Months\end{tabular}} \\ \hline
24    & \multicolumn{1}{c|}{2009}                             & Jan                            & \multicolumn{1}{c|}{2014}                            & April                           & \multicolumn{1}{c|}{82.0}                         & \multicolumn{1}{c|}{5.3}                                 & 64                                \\ \hline
\end{tabular}
\end{table}
\begin{table*}
\small
\centering
\caption{Main propagation parameters of DHO/23.40 kHz radio signal from transmitter along path to receiver at Belgrade, for 20 January 2009 at 06:16 UT.}
\label{tab:3}
\begin{tabular}{|l|l|l|l|l|l|l|l|}
\hline
   & $D$(km) & $\sigma$(Sm$^{-1}$) & $\chi$(deg) & $\beta$(km$^{-1}$) & $H'$(km) & $Ne$(m$^{-3}$) & \textit{mds} \\ \hline
1  & 0     & 0.01  & -100.9 & 0.43    & 87     & 3.09E7  & 18  \\ \hline
2  & 240   & 0.01  & -98.9  & 0.41    & 84.8   & 4.31E7  & 17  \\ \hline
3  & 460   & 1E-3  & -97.1  & 0.39    & 82.7   & 5.78E7  & 16  \\ \hline
4  & 560   & 0.01  & -96.2  & 0.39    & 82.7   & 5.78E7  & 16  \\ \hline
5  & 600   & 1E-3  & -95.9  & 0.39    & 82.7   & 5.78E7  & 16  \\ \hline
6  & 660   & 1E-3  & -95.4  & 0.37    & 80.5   & 8.08E7  & 15  \\ \hline
7  & 820   & 0.01  & -94    & 0.37    & 80.5   & 8.08E7  & 15  \\ \hline
8  & 880   & 0.01  & -93.5  & 0.34    & 78.3   & 1.13E8  & 13  \\ \hline
9  & 1100  & 0.01  & -91.7  & 0.32    & 76.2   & 1.56E8  & 10  \\ \hline
10 & 1300  & 0.01  & -90    & 0.30     & 74     & 2.18E8  & 7   \\ \hline
\end{tabular}
\end{table*}
\begin{figure*}
\begin{center}
\includegraphics[width=0.45\textwidth]{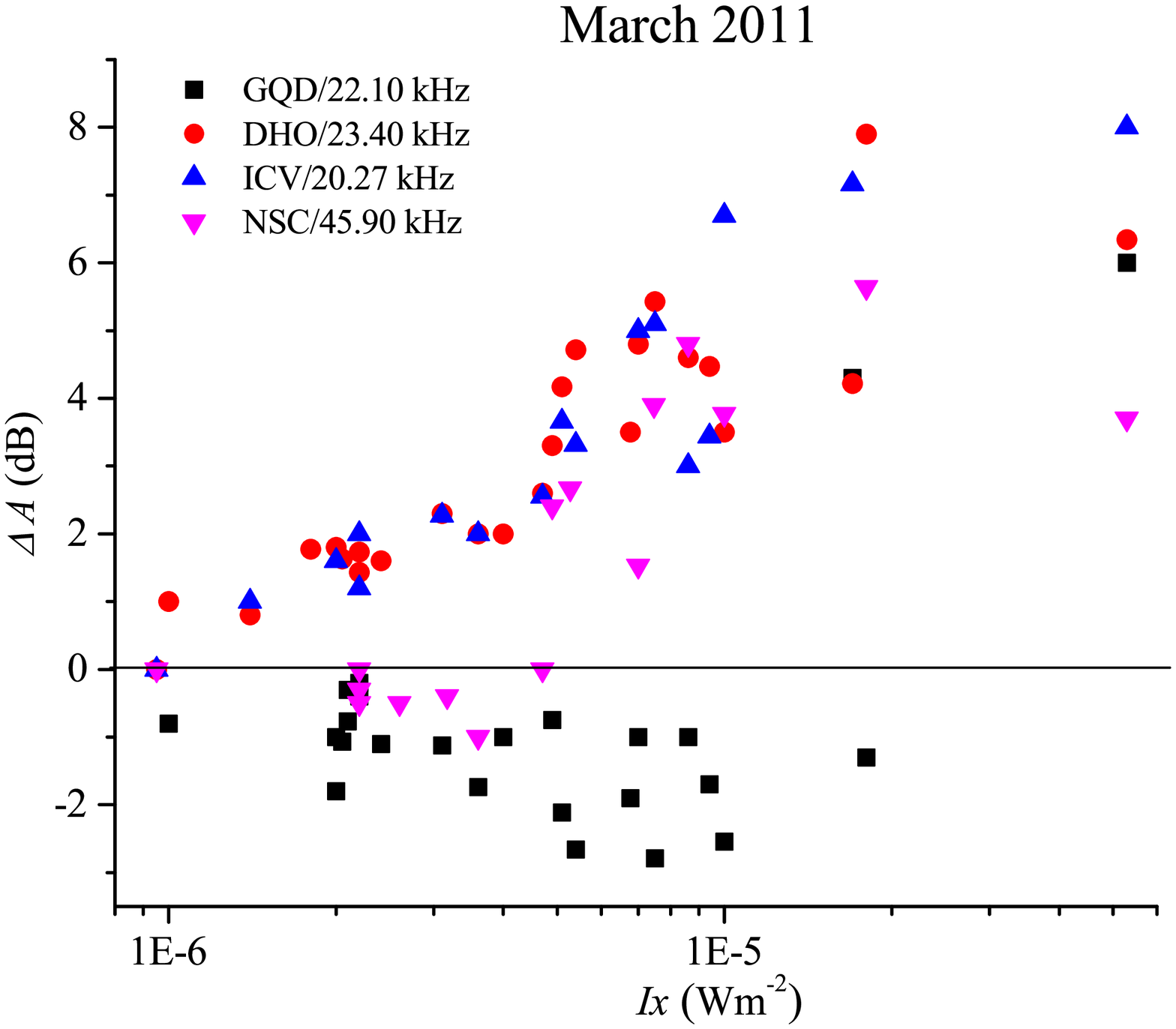}
\includegraphics[width=0.45\textwidth]{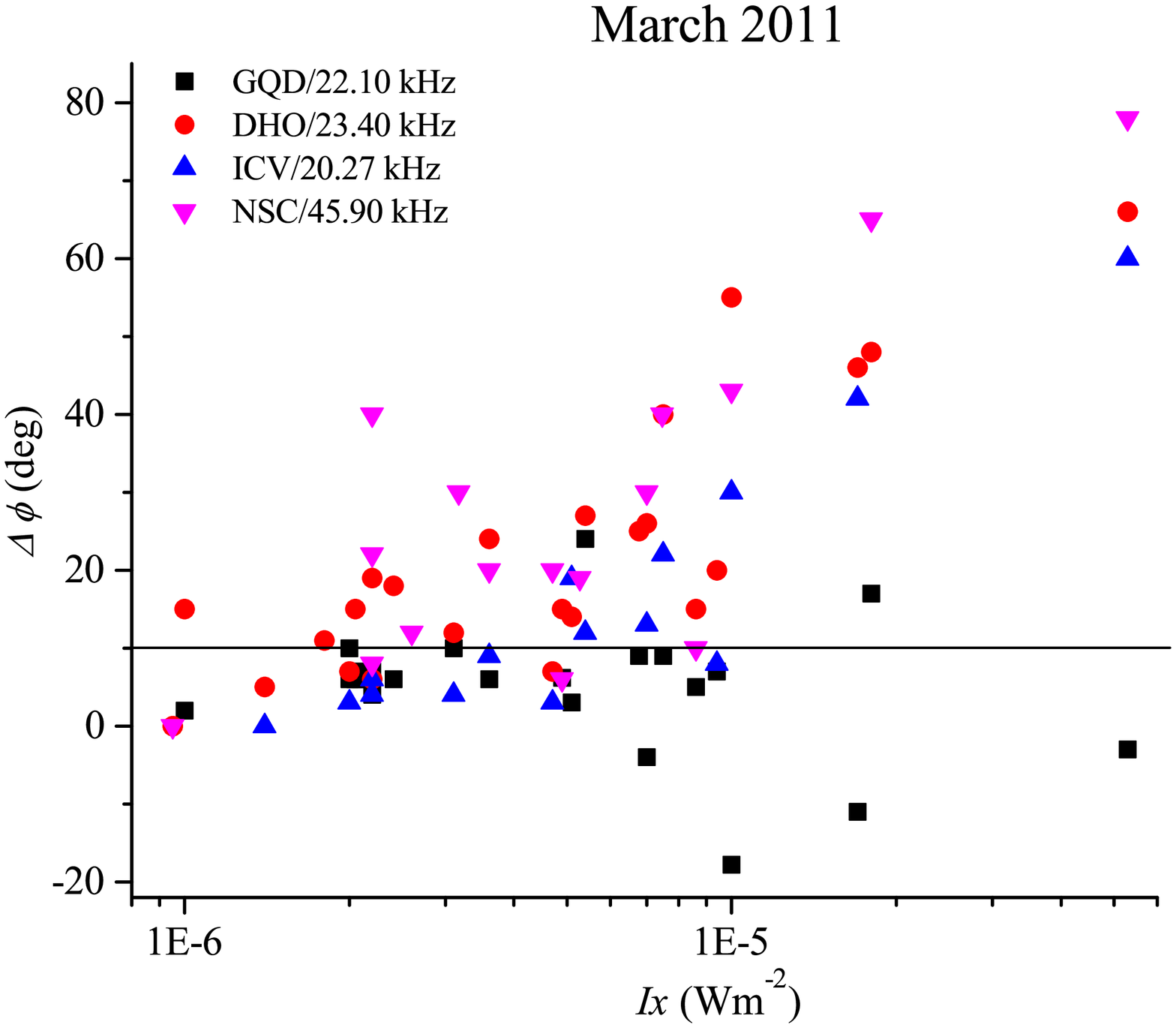}
\caption{\hspace{2in} a) \hspace{2in} b) \newline  a) Observed amplitude perturbations on GQD/22.10 kHz, DHO/23.40 kHz, ICV/20.27 KHz, and NSC/45.90 kHz radio signals as a function of X-ray flux  b) Observed phase perturbations on the same radio signals as a function of X-ray flux, measured at Belgrade during March 2011.}
\label{fig:8}
\end{center}
\end{figure*}

\subsection{Path dependent} For studying solar flare events at similar conditions we have examined only those events that occurred in the time sector for solar zenith angle $-60^{0} \le $ $\chi $ $\le 60^{0}$. As for the above mentioned facts, we have already examined 250 solar flare events in period 2009-2014, and analyzed their effects on propagation characteristics of VLF/LF radio signals.

Behaviors of amplitude and phase perturbations on VLF/ LF radio signals induced by different intensity of solar X-ray fluxes, observed on Belgrade site, were examined and these results are presented in recent papers: \citet{gru05,gru08,zig07,nin12,sul14}.
Also we studied results about the modeling of VLF wave propagation over long paths ($D$ $<$  12 Mm)  under normal and disturbed ionospheric conditions \citep{mcr00,dev08,tho11,cha12,bas13,kum14}.  As our work is based on monitoring and modeling VLF/LF wave propagation over short paths we studied results obtained by \citet{tod07,gru08,sch13,kol14}. In the next subsection are given our results for used  Wait's parameters in the RANGE model of the LWPC code, compared with the published  results.

Simultaneous observations of amplitude ($A$) and phase ($\phi$) on VLF/LF radio signals during solar X-ray flares could be applied for calculations of electron density profile. Therefore, the amplitude perturbations were estimated as a difference between values of the perturbed amplitude induced by X-ray flare and amplitude in the normal ionospheric condition: $\Delta A$ = $A_{per}$ - $A_{nor}$. Phase perturbations were estimated as: $\Delta \phi$ = $\phi_{per}$ - $\phi_{nor}$.

Although the solar X-ray flare effects on the propagating VLF/LF radio signals are well recognized on all paths, similarities and differences between them are defined under existing ionospheric conditions over the paths. According to that we will present our results of amplitude and phase perturbations on VLF radio signals observed during March 2011.

Twenty three solar X-ray flares occurred on 6 March 2011, which is unusually large number for medium solar activity. Figure \ref{fig:7}a gives an instructive example of four successive solar flares that induced amplitude perturbations on GQD/22.10 kHz, DHO/23.40 kHz, ICV/20.27 kHz and NSC/45.90 kHz radio signals observed in Belgrade site. In Figure \ref{fig:7}b there are phase perturbations on same four VLF/LF radio signals versus universal time from 08:45 UT to 12:45 UT. The top panels of Figure \ref{fig:7}a and \ref{fig:7}b show the solar X-ray flux against universal time as monitored by GOES-15 satellite.

Table \ref{tab:5} provides measured values of amplitude and phase perturbations on GQD-BEL, DHO-BEL, ICV-BEL and NSC-BEL paths induced by C2.65 ($2.65\cdot 10^{-6}$ Wm$^{-2}$ of X-ray flux in the band 0.1-0.8 nm), C3.13 ($3.13\cdot 10^{-6}$ Wm$^{-2}$), C7.53 ($7.53\cdot 10^{-6}$ Wm$^{-2}$), and C5.14 ($5.14\cdot 10^{-6}$ Wm$^{-2}$) class solar X-ray flares.

Based on the observed and presented data on Figure \ref{fig:7}a and \ref{fig:7}b it is easily to summarize the following:

\begin{enumerate}

\item  On GQD-BEL path amplitude decreases during occurrences of four C class solar flares (Figure \ref{fig:7}a, panel number 2). The size of amplitude perturbations is proportional to the intensity of solar X-ray flux. A phase change on this path is complicated, displaying increase and also decrease during the occurrence of C7.53 ($7.53\cdot 10^{-6}$ Wm$^{-2}$) class solar X-ray flare.

\item  On DHO-BEL path the amplitude and phase increase and the size of amplitude and phase perturbations are in correlation with the intensity of solar X-ray flux.

\item  On ICV-BEL path both amplitude and phase increase with the changes of intensity of X-ray flux. The shape of curves of amplitude variation with time for DHO-BEL and ICV-BEL paths are very similar to each other. The size of amplitude perturbations on DHO-BEL and ICV-BEL paths caused by same solar X-ray flare are similar to each other (Table \ref{tab:5}).

\item  Phase perturbations on NSC-BEL path display more sensitivity to the changes on intensity of solar X-ray flux than amplitude. During the occurrence of C2.65 ($2.65\cdot 10^{-6}$ Wm$^{-2}$) class solar X-ray flare there is no visible development of amplitude perturbation, while the phase increase is significant.

\end{enumerate}

\begin{table}[]
\small
\centering
\caption{Calculated parameters of normal and disturbed D-region and variation of signal amplitude.}
\label{tab:4}
\begin{tabular}{|l|l|l|l|}
\hline
\begin{tabular}[c]{@{}l@{}}Calculated \\ parameters\end{tabular}   & \begin{tabular}[c]{@{}l@{}}Normal \\ ionosphere\end{tabular} & \multicolumn{2}{l|}{SID} \\ \hline
$\beta$(km$^{-1}$)                                                         & 0.30                                                         & 0.42        & 0.52       \\ \hline
$H'$(km)                                                           & 74                                                           & 70.5        & 67         \\ \hline
$Ne$ (m$^{-3}$) at $H'$                                                   & $2.18\cdot10^{8}$                                                     & $3.93\cdot10^{8}$    & $6.23\cdot10^{8}$   \\ \hline
Number of modes                                                  & 7                                                            & 8           & 9          \\ \hline
\begin{tabular}[c]{@{}l@{}}Amplitude\\ dB above 1$\mu$V/m\end{tabular} & 31.5                                                         & 34.3        & 36.1       \\ \hline
\end{tabular}
\end{table}

\begin{table*}
\centering
\small
\caption{Measured values of amplitude and phase perturbations on four radio signals induced by small class solar X-ray flares}
\begin{tabular}{|p{1.0in}|p{0.5in}|p{0.4in}|p{0.4in}|p{0.4in}|p{0.4in}|p{0.4in}|p{0.4in}|p{0.3in}|p{0.4in}|} \hline
\multicolumn{2}{|p{1in}|}{Solar X-ray flare} & \multicolumn{2}{|p{0.8in}|}{GQD\newline 22.10 kHz} & \multicolumn{2}{|p{0.8in}|}{DHO\newline 23.40 kHz} & \multicolumn{2}{|p{0.8in}|}{ICV\newline 20.27 kHz} & \multicolumn{2}{|p{0.7in}|}{NSC\newline 45.90 kHz} \\ \hline \hline
class & $I_{Xmax}$\newline Wm$^{-2}$ & $\Delta A$\newline dB & $\Delta \phi$\newline deg & $\Delta A$\newline dB & $\Delta \phi$\newline deg & $\Delta A$\newline dB & $\Delta \phi$ \newline deg & $\Delta A$ \newline dB & $\Delta \phi$\newline deg \\ \hline
C2.65(09:04 UT) & 2.65E-6 & -0.4 & 7 & 1.43 & 6 & 1.2 & 2 & 0 & 25 \\ \hline
C3.13(10:01 UT) & 3.13E-6 & -1.4 & 10 & 2.3 & 12 & 2.27 & 6 & -0.4 & 27 \\ \hline
C7.53(10:40 UT) & 7.53E-6 & -4 & 10 & 5.43 & 48 & 5.1 & 22 & 3.9 & 40 \\ \hline
C5.14(12:09 UT) & 5.14E-6 & -2.6 & 8 & 4.17 & 26 & 3.36 & 19 & 2.67 & 19 \\ \hline
\end{tabular}
\label{tab:5}
\end{table*}

During March 2011 we examined thirty events of solar X-ray flares that occurred in the time sector for solar zenith angle $-60^{0} \le $ $\chi $ $\le 60^{0}$. Intensity of those X-ray flares was in range from C1 to M5.3 ($1\cdot 10^{-6}$ to $5.3\times10^{-5}$ Wm$^{-2}$) class. For each flare we measured amplitude and phase perturbations on four VLF/LF paths, except when the transmitter was off-air. In Figure \ref{fig:8}a and \ref{fig:8}b there are data points of the observed GQD/22.10 kHz, DHO/23.40 kHz, ICV/20.27 kHz and NSC/45.90 kHz amplitude and phase perturbations as a function of  solar X-ray flare intensity, respectively.

The range of size in amplitude and phase perturbations varies for different paths. Amplitude changes on DHO/23.40 and ICV/20.27 kHz radio signals have strong preferences to increase. In all analyzed events the phase perturbations on DHO/23.40 and ICV/20.27 kHz radio signals show an increase. The amplitude perturbations of GQD/22.10 kHz and NSC/45.90 kHz radio signals are distributed between increase and decrease, or between enhancement and attenuation, which depends on intensity of solar X-ray flux. The obtained results reveal that the phase perturbations on GQD/22.10 kHz show mainly increase and in some events decrease. The phase perturbations on NSC/45.90 kHz have a strong preference for phase increase.

We used the LWPC code for determining electron density enhancements in the D-region which were caused by flares C1 to M5.6 ($1\cdot10^{-6}$ to $5.6\cdot10^{-5}$ Wm$^{-2}$) classes that were occurred in March 2011.
The unperturbed (averaged) values of $\beta$ = 0.30 km$^{-1}$ and $H'$ = 74 km are used along four short paths, $D < 2$ Mm and calculated the initial electron density is $Ne$(74 km) = $2.18\cdot10^{8}$ m$^{-3}$.

\begin{figure}[]
\begin{center}
\includegraphics[width=\columnwidth,height=0.7\columnwidth]{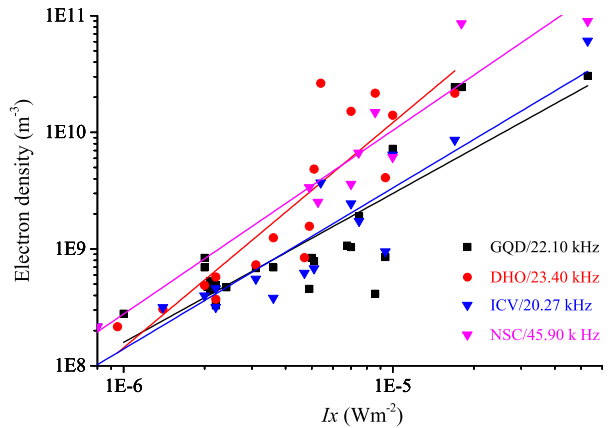} \caption{Values of electron density at height $h$ = 74 km during flare occurrences, against maximum intensity of X-ray flux calculated on the basis of VLF/LF propagation data recorded at Belgrade. Initial electron density $Ne$(74 km) = $2.18\cdot10^{8}$ m$^{-3}$.}
\label{fig:9}
\end{center}
\end{figure}

The basis for modeling altitude profile of electron density during each SID are  measured perturbations of amplitude, $\Delta A$ and phase, $\Delta \phi$ on four VLF/LF radio signals recorded at Belgrade site. The electron densities (Eq.~\ref{eq:Ne}) at height $h=74$ km for the presently obtained ($\beta_{per}$, $H'_{per}$) as a function of X-ray flare irradiance are shown in Figure \ref{fig:9}. Calculated electron density for each VLF/LF path is given by different symbol and color. For each waveguide resulting data of electron density are performed linear fitting and line is shown on Figure \ref{fig:9}. It is evident that electron density is nearly proportional to the logarithm of the X-ray irradiance maximum. Four sets of electron density display the well known increasing trend with increasing X-ray irradiance \citep{mit74}.

Our results show good correlation of linear fitting lines calculated for enhancements of electron density induced by different intensities of solar X-ray flares on GQD-BEL, DHO-BEL, ICV-BEL and NSC-BEL short paths during March 2011. These short paths have their own specifications but under major increase in the flux of X-ray, enhancements of the electron density in the D-region over Central Europe are similarly spread inducing close characteristics of the upper boundary of these waveguides.

\subsection{Distribution of amplitude perturbations during different solar activity} In this section we will present our results on amplitude and phase perturbations on DHO/23.40 kHz radio signal during different solar activity, at low, medium and high activity of current solar cycle 24. In this period, receiving DHO/23.40 kHz radio signal is stable and sensitive on increase of intensity of solar X-ray flux so recorded data are very useful for further study.

The first analyzed SID effect was induced by B8.8 ($8.8\cdot10^{-7}$ Wm$^{-2}$) class solar flare occurred on 3 November 2008, which induced amplitude and phase perturbations $\Delta A$ = 1.5 dB, $\Delta \phi = 7^{0}$, respectively. Electron density at height $h$ = 74 km changed from $Ne = 2.18\cdot10^{8} \textrm{m}^{-3}$ to $Ne = 4.56\cdot10^{8} \textrm{m}^{-3}$. From 3 November 2008 to the end of 2014 we selected 251 events of SIDs induced by different classes of solar flares.

\begin{figure*}
\begin{center}
\includegraphics[width=0.45\textwidth]{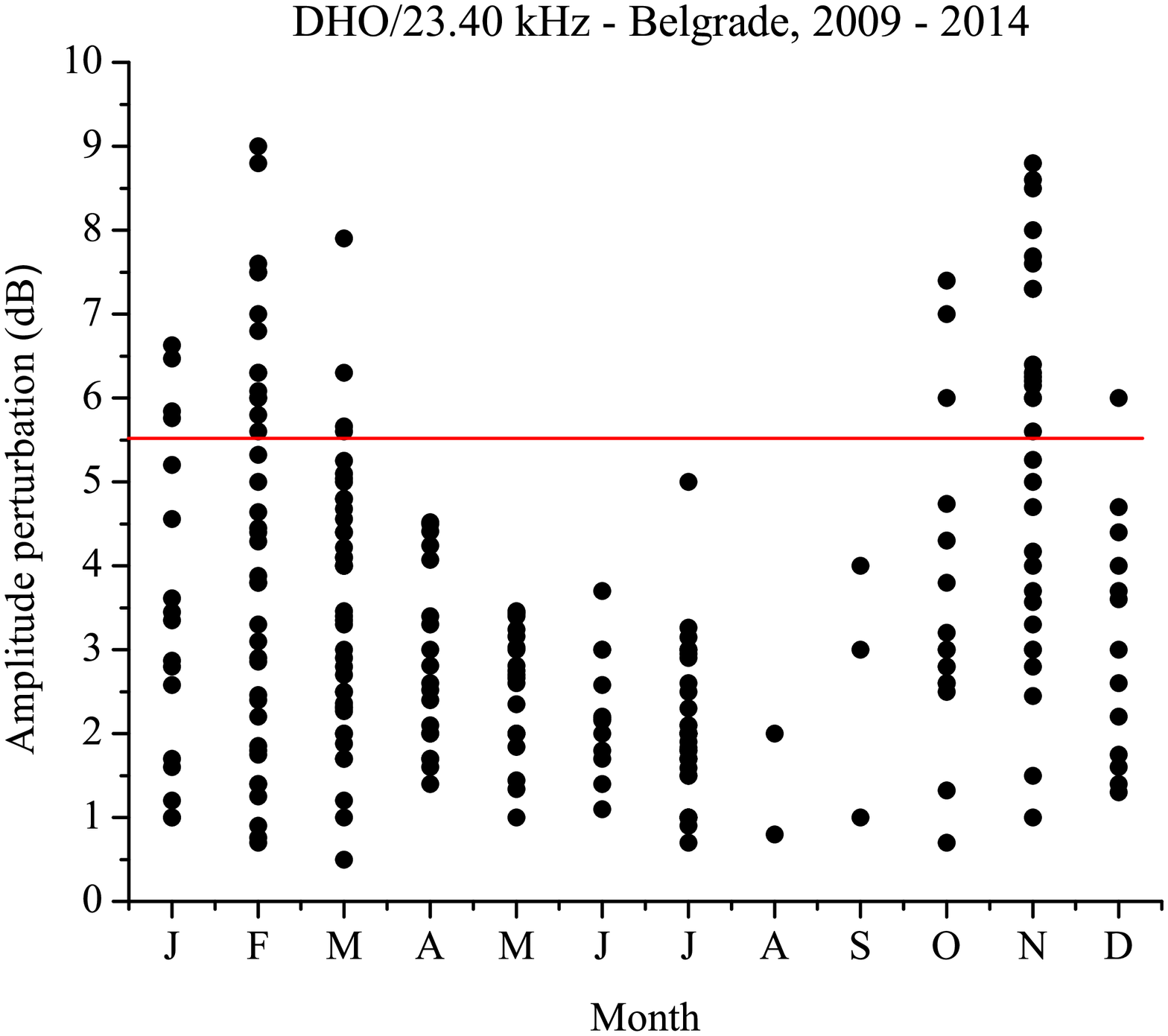}
\includegraphics[width=0.45\textwidth]{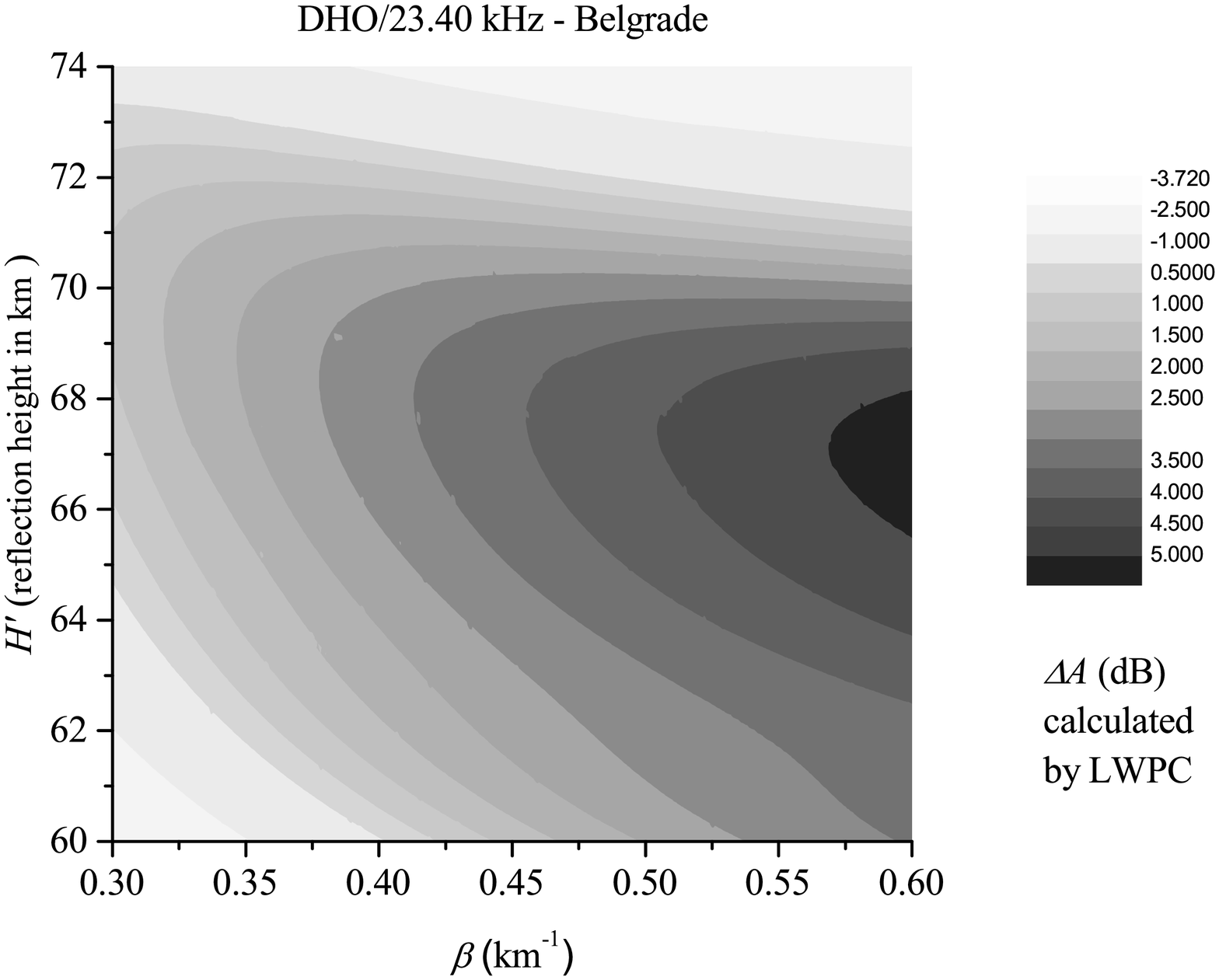}
\caption{\hspace{2in} a) \hspace{2in} b) \newline  a) Distribution of amplitude perturbations classified in accordance to the month, during 2009-2014. There is a red horizontal line that separates SIDs in two groups in accordance of amplitude perturbations. b) Counter plot of calculated amplitude perturbations as a function of Wait's parameters $\beta$ and $H'$ by the LWPC code.}
\label{fig:10}
\end{center}
\end{figure*}

Figure \ref{fig:10}a shows the distribution of amplitude perturbations on DHO/23.40 kHz classified in accordance to the months over year. The size of amplitude perturbations is in the range 0.5 dB $< \Delta A <$ 9 dB. In Figure \ref{fig:10}a there is a difference in values of amplitude perturbations recorded during winter and summer months.

The observed amplitude and phase perturbations of DHO/23.40 kHz were simulated by the LWPC code. \citet{mcr04} found that  the increase of electron density due to a flare of X5 ($5\cdot10^{-4}$ Wm$^{-2}$) class lowers $H'$ from mid-day value of $\sim$ 71 km down to $\sim$ 58 km and increases $\beta$ from 0.39 km$^{-1}$ to 0.52 km$^{-1}$. \citet{gru08} from recorded data of GQD/22.10 kHz radio signal at Belgrade found that a series of flares from C1 to M5 ($1\cdot10^{-6}$ to $5\cdot10^{-5}$ Wm$^{-2}$) lower $H'$ from 74 km to 63 km and change $\beta$ in range 0.30-0.49 km$^{-1}$. On the basis of these results we changed $\beta$ in steps of 0.01 km$^{-1}$ and $H'$ in steps of 0.1 km as input parameters in RANGE model of the LWPC code. Calculated amplitude perturbations as a function of $H'$ (range 60-74 km) and $\beta$ (range 0.30-0.60 km$^{-1}$) are given in the counter plot for the daytime conditions at DHO/23.40 kHz radio signal in Figure \ref{fig:10}b.

Theoretically results show that maximum of amplitude perturbation is $\Delta A \sim 5.5$ dB, which is lower than measured amplitude perturbations up to $\Delta A$ $<$ 9 dB.
Similar results to ours about measured amplitude perturbations are published by \citet{tod07}. Authors successfully used the \underline{F}inite \underline{D}ifference \underline{F}requency \underline{D}omain (FDFD) model to obtain electron density in the D-region on the basis of propagation VLF radio signals (Japanese transmitters) over short paths.

With the restriction of $\Delta A < $  5.5 dB that were monitored on DHO/23.40 radio signal in period 2011-2014, we selected 201 SIDs for further study. The main aim of our work is to calculate the enhancement of electron densities caused by small and moderate classes of solar X-ray flares.

\begin{figure*}
\begin{center}
\includegraphics[width=0.47\textwidth]{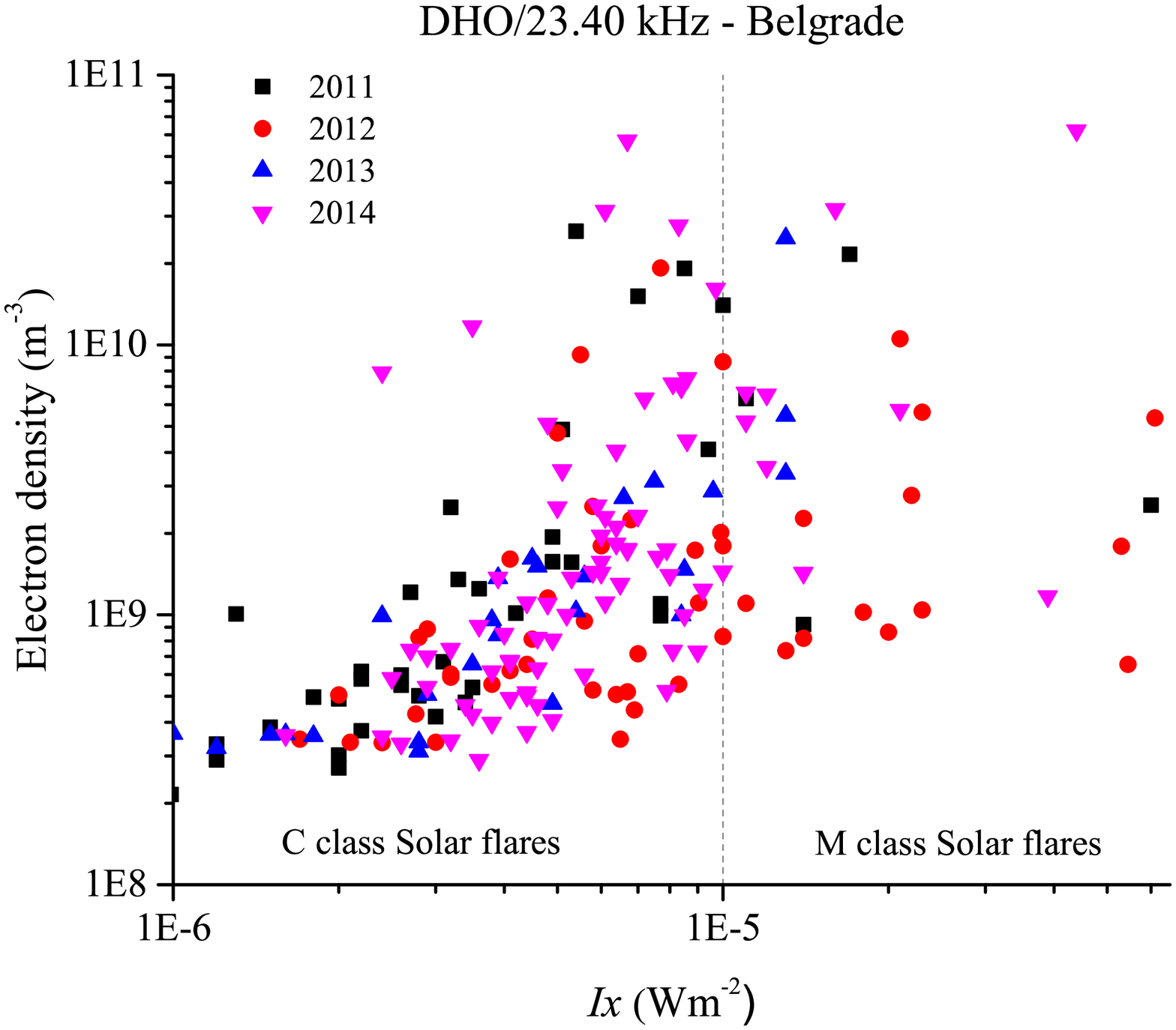}
\includegraphics[width=0.47\textwidth]{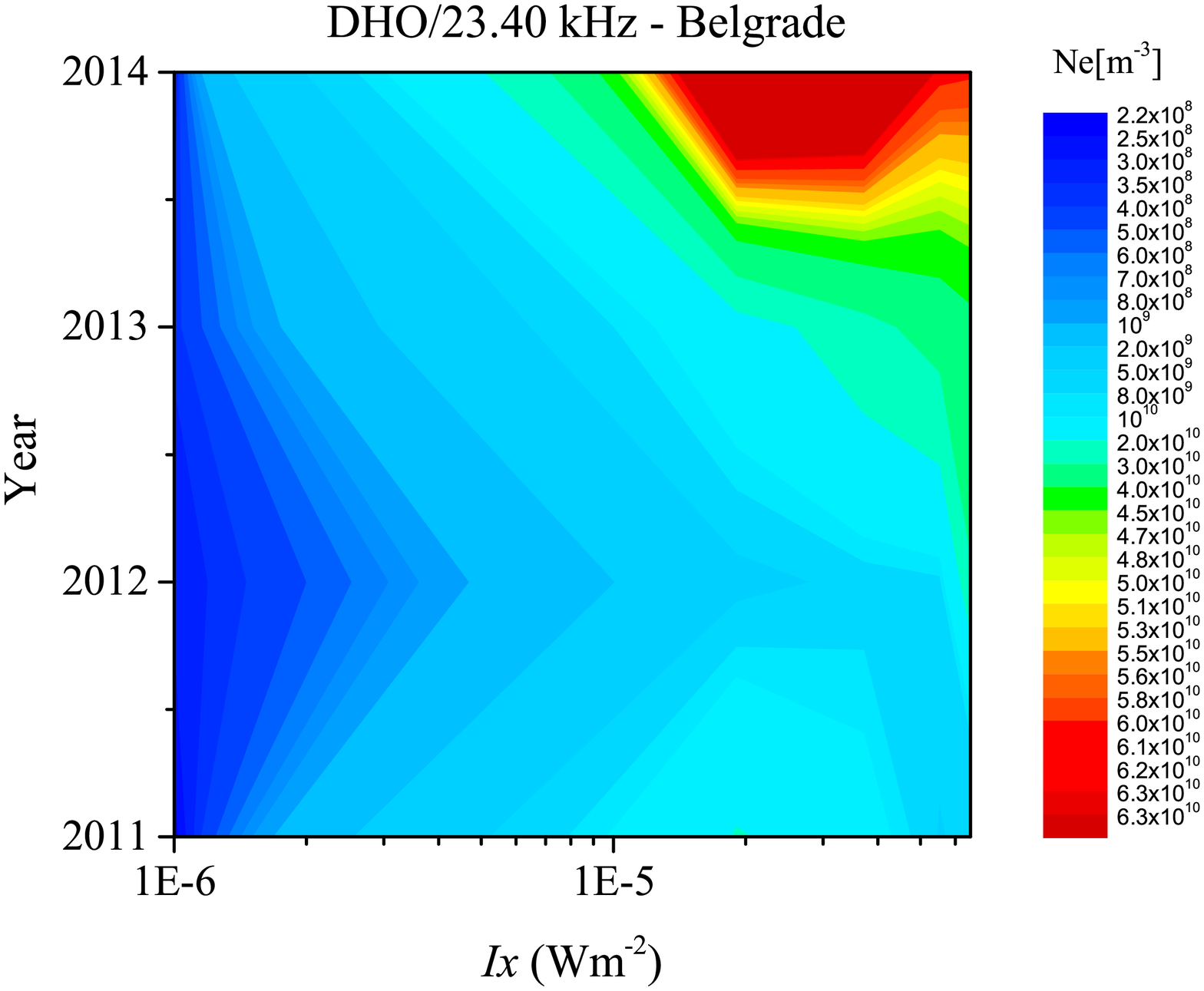}
\caption{\hspace{1.6in} a) \hspace{2.3in} b) \newline Electron density at reference height $h$ = 74 km as a function of X-ray flux during SIDs in period 2011-2014 b) surface plot of electron density as a function of X-ray flux during SIDs.}
\label{fig:11}
\end{center}
\end{figure*}

Figure \ref{fig:11}a shows calculated electron densities at height $h$ = 74 km as a function of X-ray flare intensity, obtained on basis of DHO/23.40 kHz data recorded at Belgrade site. Different colors present the calculated electron densities for each year in period  2011-2014.  For taking a view of changing electron density as a function of intensity of X-ray flux we presented our results as a surface plot, Figure \ref{fig:11}b. Generally, the results show that solar X-ray flares:

\begin{enumerate}
\item  from C1 to C6 ($1\cdot 10^{-6}$ Wm$^{-2}$ to $6\cdot 10^{-6}$ Wm$^{-2}$), class induced enhancement of electron density is from $Ne$ $\sim 2\cdot 10^{8}$ m$^{-3}$ to $Ne$ $\sim 2\cdot 10^{9}$ m$^{-3}$, respectively

\item  from C7 to M1 ($7\cdot 10^{-6}$ Wm$^{-2}$ to $1\cdot 10^{-5}$ Wm$^{-2}$) class induced electron density rises up to values $Ne$ $\sim 10^{10}$ m$^{-3}$,

\item  from M1 to M6.1 ($1\cdot 10^{-5}$ Wm$^{-2}$ to $6.1\cdot10^{-5}$ Wm$^{-2}$) class, the enhancement of electron density is different according to level of solar activity and rises up to values $Ne$ $\sim 6\cdot 10^{10}$ m$^{-3}$,

\item  the greatest enhancement of electron densities (red color) are induced by moderate class solar X-ray flares, which occur more frequently in the year of the maximum of solar cycle 24.
\end{enumerate}

\section{Summary and conclusions}

The purpose of this work was to analyze the amplitude and phase data acquired by monitoring at Belgrade site VLF/LF radio signals emitted by four European transmitters during a seven-year period (2008-2014). The results of amplitude and phase variations on GQD/22.10 kHz, DHO/23.40 kHz, ICV/20.27 kHz and NSC/45.90 kHz radio signals measurements at short path over Central Europe and their interpretation are summarized here. The most important factors affecting those paths under uniform background conditions are the transmitter frequency, geographical location, the electron density profile, and the ground conductivities encountered.

Our attention is restricted to regular diurnal, seasonal and solar variations including sunrise and sunset effects on propagation characteristics of VLF/LF radio signals. All the paths are similarly illuminated during daytime condition and there are differences in the level of illumination during dawn and dusk in accordance to geographic coordinates of transmitters. We  accepted  the results obtained by \citet{vol64} that VLF radio signals  propagating  from transmitter to receiver over paths with distance $D$ $<$ 2 Mm reflect once on the middle of the paths and propagate as a superposition of discrete modes. In general, for propagation of  VLF/LF radio signals over short paths from transmitters to Belgrade site we conclude that:

\begin{itemize}
  \item[-] The diurnal variation of amplitude is caused by smooth variation of the equivalent reflection height and the reflection characteristics of the D-region during 24 hours.

  \item[-] On the basis of changing reflection characteristics of the D-region our numerical results show that propagation of VLF radio signal is created as a superposition of $n_{n}$ $\sim$ 17 discrete modes during nighttime and $n_{d}$ = 7 during daytime condition. Propagation of LF radio signal is performed with $n_{n}$ = 34 (nighttime) and $n_{d}$ = 10 (daytime) discrete modes. Sunrise effects on VLF/LF propagating over a short path cause a gradual fall of number of discrete modes. This implicates that the number of the discrete modes is induced by the transmitted frequency.

  \item[-] The process of ionization in the D-region begins when solar zenith angle has value $\chi = -99^0$, and sunrise terminator reaches the height $h = 95$ km. When this process starts in the middle of the propagation path, the consequence is the development of the first amplitude minimum and the transition from phase level during nighttime to phase level during daytime.

  \item[-] Based on specifications for each path and a function of diurnal-seasonal variation, signal characteristics morphology, i.e. how many amplitude minima and at what time they would be developed is defined. It is characterized by periodic and repeatable variations of amplitude as the dawn-dusk terminator moves along a VLF/LF path.

  \item[-] On the basis of measured VLF/LF data our conclusion is that the appearances of amplitude minima over short path could be divided in \textit{two types}. The amplitude minima that are appeared in time intervals during transition of nighttime/daytime and daytime/nighttime conditions on the middle of the propagation path belong to the first type. The amplitude minima that occur under daytime condition over all short paths belong to the second type. Usually they are developed as a pair and their timings are symmetrical arrange in a according to a local noon. Timings of their occurrences continuously change from day to day.
\end{itemize}

Although the solar X-ray flare effects on propagation of VLF/LF radio signals are well recognized on all paths, similarities and differences between them are defined under existing conditions over the paths. Results about propagation characteristics of signal amplitude and phase show that each short path is unique and thus each path reacts differently to the electron density enhancement induced by same solar X-ray flare. As it can be seen from this study the range of size in amplitude and phase perturbations varies for different paths and also statistical results show that the size of amplitude and phase perturbations on VLF/LF radio signal is in correlation with the intensity of X-ray flux. The main results are:

\begin{itemize}
  \item[-] The DHO/23.40 kHz and the ICV/20.27 kHz radio signals measured at Belgrade site always show amplitude increases and strong preference for phase increases under enhancement of electron density induced by solar X-ray flare.
  \item[-] Amplitude changes of GQD/22.10 kHz radio signal are evenly distributed between enhancement and attenuation. Phase changes show a different character, displaying increase and also decrease during the occurrence of different solar flare classes.
  \item[-] The NSC/45.90 kHz radio signal at Belgrade site always shows more sensitivity of phase perturbation to the changes on intensity of solar X-ray irradiance than amplitude.
\end{itemize}

The model computations applied to the amplitude and phase perturbations on VLF/LF radio signals recorded at Belgrade site are able to reproduce the general features of the electron density enhancement induced by occurrence of solar X-ray flares during the period of ascending phase and maximum of the solar cycle 24.

\section*{Acknowledgments}
The authors are thankful to the Ministry of Education, Science and Technological Development of the Republic of Serbia for support of this work within projects 176002 and III44002. The authors also express their thanks to the reviewers for their comments and valuable suggestions.

\bibliographystyle{elsarticle-harv}

\end{document}